\documentclass[trackchanges,twocolumn,twocolappendix,longbib]{aastex701}
\hypersetup{linkcolor=red,citecolor=green,filecolor=cyan,urlcolor=magenta}
\usepackage{amsmath}

\begin{document}

\title{3D MHD simulations of coronal loops heated via magnetic braiding \\ II. Automatic detection of reconnection outflows and statistical analysis of their properties}

\author[orcid=0009-0005-4545-1010,sname='Gabriele Cozzo']{G. Cozzo}
\affiliation{Harvard–Smithsonian Center for Astrophysics, 60 Garden St., Cambridge, MA 02193, USA}
\affiliation{INAF-Osservatorio Astronomico di Palermo, Piazza del Parlamento 1, I-90134 Palermo, Italy}
\email[show]{gabriele.cozzo@cfa@harvard.edu}  

\author[orcid=0000-0002-0405-0668, sname='Paola Testa']{Paola Testa} 
\affiliation{Harvard–Smithsonian Center for Astrophysics, 60 Garden St., Cambridge, MA 02193, USA}
\email{ptesta@cfa.harvard.edu}

\author[orcid=0000-0002-0333-5717, sname='Juan Martinez-Sykora']{J. Martinez-Sykora}
\affiliation{Lockheed Martin Solar and Astrophysics Laboratory, 3251 Hanover St, Palo Alto, CA 94304, USA}
\affiliation{Rosseland Centre for Solar Physics, University of Oslo, P.O. Box 1029 Blindern, N-0315 Oslo, Norway}
\affiliation{Institute of Theoretical Astrophysics, University of Oslo, P.O. Box 1029 Blindern, N-0315 Oslo, Norway}
\affiliation{SETI Institute, 339 Bernardo Ave, Suite 200, Mountain View, CA, 94043, United States}
\email{jmsykora@seti.org} 

\author[orcid=0000-0001-5274-515X, sname='Paolo Pagano']{P. Pagano}
\affiliation{Dipartimento di Fisica \& Chimica, Università di Palermo, Piazza del Parlamento 1, I-90134 Palermo, Italy}
\affiliation{INAF-Osservatorio Astronomico di Palermo, Piazza del Parlamento 1, I-90134 Palermo, Italy}
\email{paolo.pagano@unipa.it} 

\author[orcid=0000-0002-1820-4824, sname='Fabio Reale']{F. Reale}
\affiliation{Dipartimento di Fisica \& Chimica, Università di Palermo, Piazza del Parlamento 1, I-90134 Palermo, Italy}
\affiliation{INAF-Osservatorio Astronomico di Palermo, Piazza del Parlamento 1, I-90134 Palermo, Italy}
\email{fabio.reale@unipa.it} 

\author[orcid=0000-0001-9030-0418, sname='Antonino Franco Rappazzo']{F. Rappazzo}
\affiliation{Dipartimento di Fisica \& Chimica, Università di Palermo, Piazza del Parlamento 1, I-90134 Palermo, Italy}
\affiliation{INAF-Osservatorio Astronomico di Palermo, Piazza del Parlamento 1, I-90134 Palermo, Italy}
\email{antoniofranco.rappazzo@unipa.it} 

\author[orcid=0000-0003-0975-6659, sname='Viggo Hansteen']{V. Hansteen}
\affiliation{Rosseland Centre for Solar Physics, University of Oslo, P.O. Box 1029 Blindern, N-0315 Oslo, Norway}
\affiliation{Institute of Theoretical Astrophysics, University of Oslo, P.O. Box 1029 Blindern, N-0315 Oslo, Norway}
\affiliation{SETI Institute, 339 Bernardo Ave, Suite 200, Mountain View, CA, 94043, United States}
\email{vhansteen@seti.org} 

\author[orcid=0000-0002-8370-952X, sname='Bart De Pontieu']{B. De Pontieu}
\affiliation{Lockheed Martin Solar \& Astrophysics Laboratory, 3251 Hanover St, Palo Alto, CA 94304, USA}
\affiliation{Rosseland Centre for Solar Physics, University of Oslo, P.O. Box 1029 Blindern, N-0315 Oslo, Norway}
\affiliation{Institute of Theoretical Astrophysics, University of Oslo, P.O. Box 1029 Blindern, N-0315 Oslo, Norway}
\email{bdp@lmsal.com} 

\begin{abstract}

Recent observations of fast and bursty ``nanojets'' suggest novel diagnostics of nanoflare heating in the solar corona.
The aim of this work is to investigate the presence and properties of reconnection outflows, similar to observed nanojets, in numerical simulations, and explore their relationship with the nanoflare properties. This work explores  their potential as diagnostics for nanoflare heating in observations. We developed an algorithm of Reconnection Outflows Automatic Detection (ROAD) in 3D MHD simulations of coronal loops. We applied the algorithm to a 3D MHD stratified coronal loop model heated by magnetic reconnection and analyzed the statistical properties of the jets produced at reconnection sites, over about one solar hour. The magnetic structure is maintained at high temperature and for an indefinite time by intermittent episodes of local magnetic energy release due to reconnection.

\end{abstract}


\section{Introduction}


In the non-flaring corona, the heating is thought to arise from numerous small, impulsive energy releases, the so-called ``nanoflares'' \citep{parker1988nanoflares}, generated as narrow current sheets reconnect throughout the corona \citep[see reviews by][]{klimchuk2015key, parnell2012contemporary}.  Individual nanoflares are thought to be too small and frequent to be isolated observationally with current instruments, yet EUV and X-ray detections of brief, high-temperature bursts suggest their presence \citep[e.g.,][]{reale2011solar,testa2012hinode,testa2013observing,testa2014evidence,reale2014coronal,cargill2015modelling,ishikawa2017detection, testa2020, cho2023}.  Understanding how such ubiquitous, small-scale reconnection episodes combine to sustain the million-degree corona requires high-resolution observations and fully three-dimensional numerical models.

Recent high-cadence, sub-arcsecond observations from IRIS \citep{de2014interface} and SDO/AIA \citep{pesnell2012solar,lemen2012atmospheric} have revealed a new class of small, short-lived, jet-like features, the so-called ``nanojets'' \citep{antolin2021reconnection}, that are now regarded as key diagnostics of reconnection-driven coronal heating.  These events appear as bursts of ultraviolet or EUV emission that have a duration of tens of seconds, immediately followed by collimated plasma jets that shoot sideways across the ambient magnetic field at $100\mathrm{-}300\;\mathrm{km\,s^{-1}}$ \citep{antolin2021reconnection,sukarmadji2022observations,patel2022hi}.  Such nanojets have been associated prevalently with MHD avalanches, Kelvin-Helmholtz or Rayleigh-Taylor instabilities \citep{sukarmadji2022observations}, or to catastrophic cooling episodes with widespread evidence of coronal rain \citep{antolin2015multi}, and can even excite transverse MHD waves that feed additional loop heating \citep{sukarmadji2024transverse}. Typical widths are only $500\text{–}1500\;\mathrm{km}$; lengths extend to a few $\mathrm{Mm}$. The small dimensions and short lifetime of these features make their detection often challenging. Multi-passband observations of nanojets suggest a multi-thermal structure, while frequent asymmetries in jet propagation (nanojets are observed to be mostly one-directional) suggest an origin in braided or curved field geometries, rather than in symmetric two-dimensional current sheets \citep{pagano2021modelling,patel2022hi}, or in a non-uniform ambient medium.  
Stereoscopic SDO and STEREO data also revealed many small-scale jets, brightenings, and bidirectional flows evolving from a double-decker filament formed in a magnetically braided flux rope.
Similar events were also found within high-cadence Solar Orbiter/EUI images of an erupting filament \citep{gao2025reconnection, bura2025dynamics}, with velocity exceeding $450\,\mathrm{km}\,\mathrm{s}^{-1}$, far above earlier reported. In each case, the observed nanoflare-scale energies and geometry point to component reconnection in misaligned filament threads \citep{liu2025deciphering}. 
Small-scale, jet-like structures have also been recorded in different contexts, by, e.g., \cite{chen2017solar, chen2020coronal}.

3D and 2.5D MHD simulations addressed the issue of nanojet acceleration with ad-hoc setups where field lines are tilted by photospheric motions \citep{antolin2021reconnection} or where their misalignment is provided from the outset via the initial conditions \citep{pagano2021modelling}. 
In particular, in a three-dimensional MHD model of interacting loop strands \citep{antolin2021reconnection} the morphology and energetics of bidirectional $v\,\sim\,200\,\mathrm{km\,s^{-1}}$-fast jets match the observations and offer a theoretical framework for nanojets as a consequence of the slingshot effect from the magnetically tensed, curved magnetic field lines reconnecting at small angles. 
More recently, \cite{de2022probing} and \cite{2025A&A...695A..40C} showed that reconnection outflows can naturally develop from braiding and their evolution can be detected both in hot and cold lines and analyzed with high-resolution spectroscopy. In particular, 
\cite{2025A&A...695A..40C} considered a 3D MHD model of a dynamic coronal loop environment where an MHD avalanche is occurring and analyzed a clear outflow from a reconnection episode soon after the instability.
This hot ($\sim 8\,$MK) reconnection outflow is found to share many features with nanojets recently detected at lower temperatures, such as a total energy of $10^{24}\,$erg, a velocity of a few hundred $\mathrm{km}\,\mathrm{s}^{-1}$, and a duration of less than $1\,\mathrm{min}$.
\cite{sen2025merging} describe a 2.5D MHD simulation where impulsive perturbations in a stratified coronal current sheet generate multiple plasmoids that subsequently merge.  Reconnection at the merger interface produces short-lived, collimated bi-directional jets ($\sim 100\,\mathrm{km}\,\mathrm{s}^{-1}$, $\sim 3 \times 10^{24}\,\mathrm{erg}$) with all the hallmarks of observed nanojets, providing a complementary, plasmoid-driven mechanism for nanojet formation. 

\begin{figure}[pt]
   \centering
\includegraphics[width=\hsize]{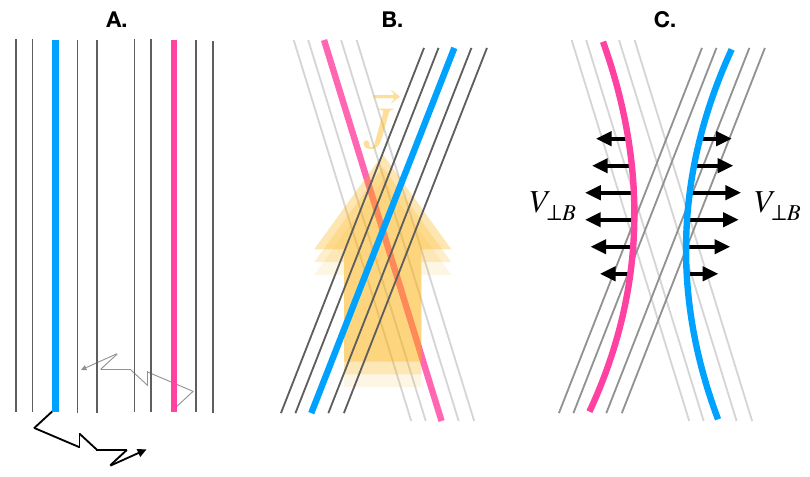}
  \caption{Schematic representation of component magnetic reconnection. Step A: Photospheric random motions drag the footpoints of two bundles of field lines.
  Step B: The two sets of field lines are tilted in opposite directions and a current sheet builds up between them. Step C: the magenta and blue field lines reconnects and accelerates two outflows in opposite direction, perpendicularly to the guide field.} 
  \label{Fig:paper_2_scheme}
\end{figure}

\begin{figure*}[ht!]
   \centering
\includegraphics[width=\hsize]{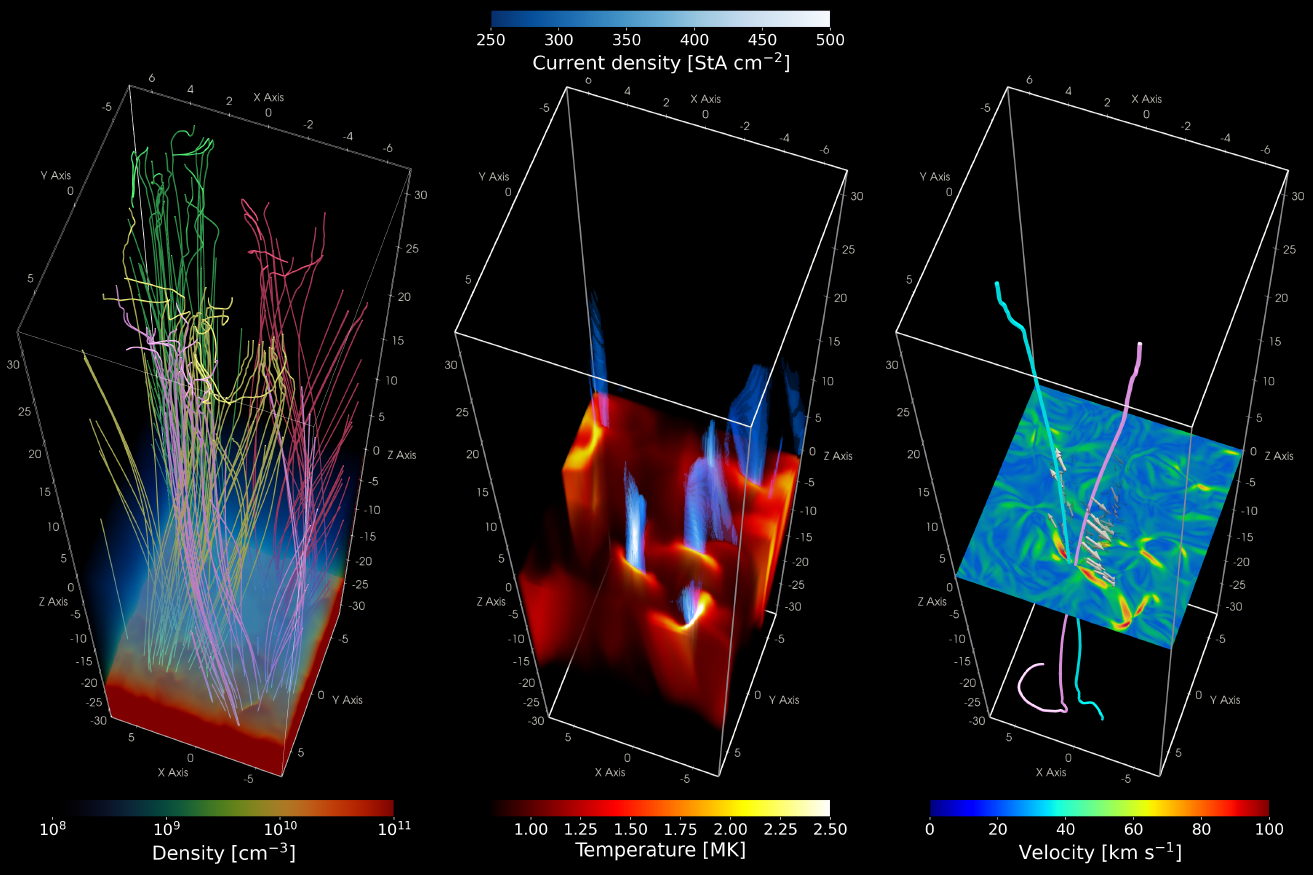}
  \caption{3D rendering of the magnetically braided coronal flux tubes. Left panel: Stratification of the plasma density at the lower footpoints from the dense chromosphere ($n > 10^{10}\,\mathrm{cm}^{-3}$, reds) to the tenuous corona ($n < 10^{9}\,\mathrm{cm}^{-3}$, blues). Four coloured bundles of field lines are shown. 
  Middle panel: The lower half of the box showcases the temperature distribution with sparse, yellow coloured, hot-spots (as high as $3\,\mathrm{MK}$) localized around narrow current sheets, the blue-to-white strips in the top half domain.
  Right panel: The mid-plane slice shows the magnitude of the velocity component perpendicular to the magnetic field. Two reconnecting field lines intersect where the velocity is high ($\sim 100\,\mathrm{km}\,\mathrm{s}^{-1}$ in the red spots). 
  Arrows show the strength and orientation of the outflow in the proximity of the magnetic field lines, similar to Fig. \ref{Fig:paper_2_scheme}. Movie I shows a $\sim 500\,\mathrm{s}$ long temporal evolution of the three panels, including field line braiding (left), progressive current sheet build-up and dissipation into heat (center), and outflows acceleration during component-reconnection (right).} 
  \label{Fig:paper_2_3D_view_1}
\end{figure*}

The next step is to assess whether observed nanojets represent systematic outcomes of component magnetic reconnection in coronal loops heated by magnetic braiding, or if they are simply serendipitous occurrences. As originally suggested by Parker \citep{parker1972topological}, MHD models demonstrate that even simple magnetic fields can produce multiple current sheets through random driving footpoint motions \citep[e.g.,][]{rappazzo2007coronal,rappazzo2008nonlinear}.
The reduced MHD experiments described by \cite{rappazzo2010shear, rappazzo2013field} concluded that the observed turbulent dynamics originate from the inherent non-linear nature of the system rather than the complexity of the photospheric driver, while \cite{ritchie2016dependence} showed that coherent footpoint motions typically lead to fewer large energy-release events, whereas complex motions result in more frequent but less energetic heating events. \cite{rappazzo2017coronal} examined magnetic topology and random walk of reconnecting field lines in a magnetically confined, nanoflare-heated corona. They are shown to form coherent structures around current sheets named ``current sheet connected regions''. 
In the accompanying paper, hereafter referred to as ``Paper I'', we presented a Pluto \citep{mignone2007pluto} 3D MHD model of a multi-strand flux tube in a stratified solar atmosphere, driven by photospheric twisting at the boundaries. We described how the dissipation of narrow current sheets leads to widespread heating, and showed that the magnetic structure can be indefinitely maintained at coronal temperatures through intermittent episodes of local magnetic energy release caused by reconnection.
In general, the evolution of plasma and magnetic field in 3D MHD models of DC heating \citep[e.g.,][and Paper I]{johnston2025self, breu2022solar, howson2022effects, reid2020coronal, reale20163d} is inherently complex and thus, it is very difficult to detect and analyze the nature of the individual heating events. As discussed above, observations suggest that impulsive outflows may also be a fair indicator of the occurrence of magnetic reconnection in braided magnetic structures. We therefore aim to investigate reconnection outflows as the aftermath of component magnetic reconnection in braided magnetic structures and their relationship with coronal heating.

The forthcoming coronal spectrometer MUSE \citep{de2020multi} will deliver maps of Doppler shifts and non-thermal velocities at unprecedented spatial and temporal resolution, enabling the detection of fast reconnection outflows and thus providing a fundamentally new diagnostic of impulsive coronal heating \citep{de2022probing}. 
Establishing a clear link between component-reconnection heating processes and the associated fast outflows is now needed to build a solid theoretical framework for the interpretation of new forthcoming observations.

In this perspective, we devised a tool for Reconnection Outflows Automatic Detection (ROAD) in 3D MHD numerical simulations of the corona, described in Section \ref{sec:methods}. 
As a proof of concept, we have applied the algorithm to the 3D MHD simulation described in Paper I, and here summarized in Section \ref{sec:model}.
We characterised the morphology, kinematics, and energetics of the detected events, and quantified their efficiency in converting magnetic energy into heat. 
We present the results of the analysis, including individual and statistical properties of the detected outflows in Section \ref{sec:results} and discuss them in Section \ref{sec:conclusions}.

\section{The Model}
\label{sec:model}

\begin{figure*}[ht!]
   \centering
\includegraphics[width=\hsize]{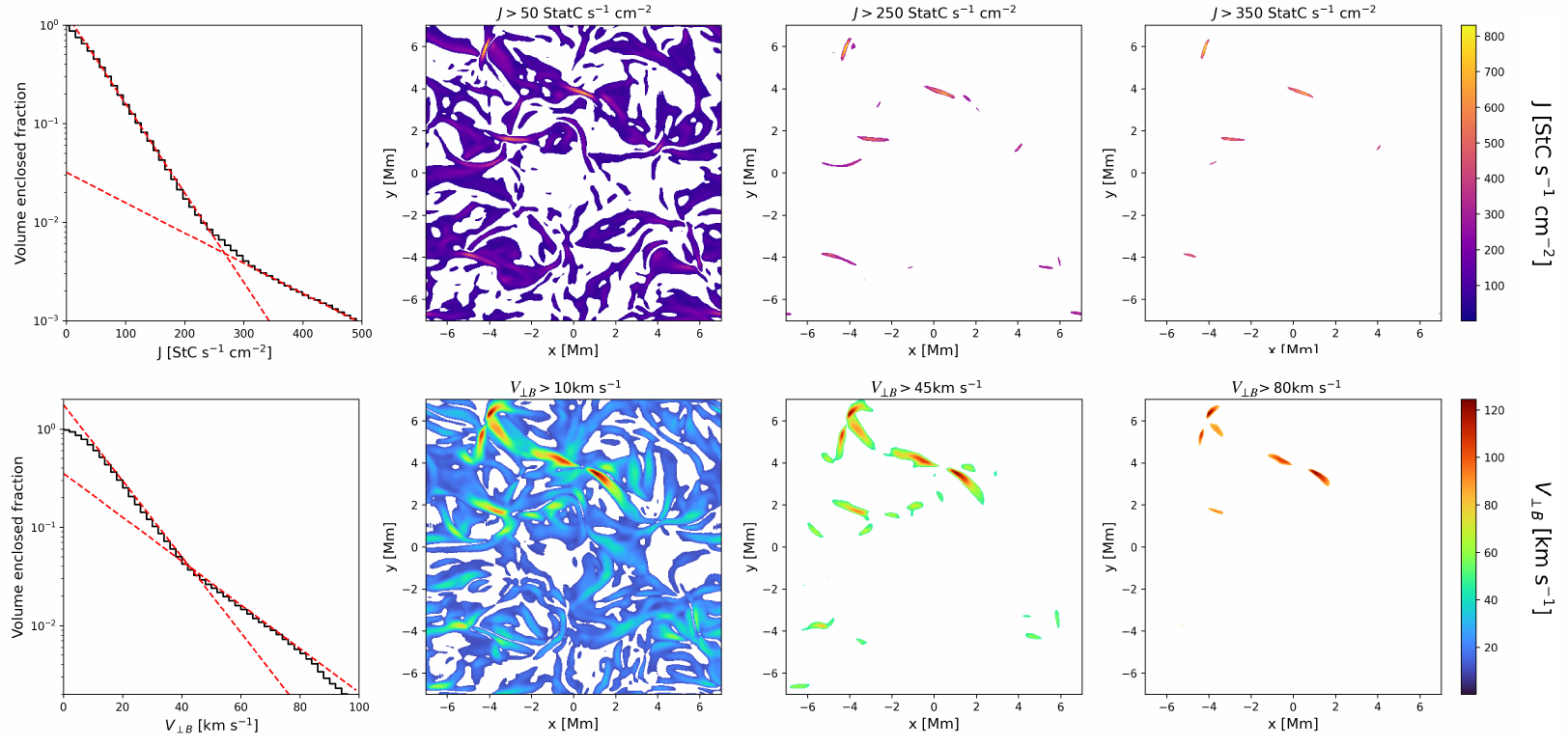}
  \caption{Cumulative distribution of the plasma current density magnitude (upper panels) and velocity component perpendicular to $\vec B$ (lower panels). Left plot: fraction of the volume enclosed by plasma with current density (velocity) up to $J^{tr}$ ($V_{\perp B}^{tr}$). The best-fit functions  above and below $J^{tr} \sim 250\,\mathrm{StC}\,\mathrm{s}^{-1}\,\mathrm{cm}^{-2}$ ($V_{\perp B}^{tr} \sim 45\,\mathrm{km}\,\mathrm{s}^{-1}$) are indicated (red dashed lines). Right panels: Maps of the current density (velocity $\perp B$) magnitude at the mid-plane. The regions with $J < J_{\perp}^{tr}$ ($V < V_{\perp}^{tr}$) are masked out.} 
  \label{Fig:paper_2_velocity_treshold}
\end{figure*}

We consider the three–dimensional MHD simulation presented in Paper I, which describes an MHD avalanche triggered in a kink-unstable, multi-threaded coronal loop system \citep[see][]{2016mhhoodd,reid2018coronal,reid2020coronal,cozzo2023coronal,cozzo2024coronal}.
Four identical magnetic flux tubes are embedded in a gravitationally-stratified atmosphere consisting of a $\sim 10^{4}\,$K, isothermal chromospheric layer at both footpoints and a $\sim 1\,$MK corona above it. Each flux tube is rooted in the chromosphere, with a magnetic field strength of $200\,G$ that reduces to $20\,\mathrm{G}$ as it expands with height in the corona.  A continuous twisting by oppositely directed boundary photospheric vortices is imposed as steady rotational motion at the footpoints of each flux tube \citep{reale20163d,cozzo2023asymmetric}.

As the imposed twisting approaches the kink threshold, each tube becomes unstable and fragments into a turbulent network of current sheets that undergo rapid, small-angle reconnection. The avalanche propagates from the first initially unstable strand to its neighbours, converting magnetic energy into heat and driving local temperature spikes up to $\sim 10\,$MK. After the global disruption, which was discussed in detail in \citet{cozzo2023coronal, cozzo2024coronal}, the steady footpoint driving continues to inject magnetic energy, perpetually regenerating current sheets within the already braided field. The system settles into a statistical balance in which Ohmic dissipation, thermal conduction, and radiative cooling approximately offset one another, yet no true steady state is reached: the corona experiences an ongoing storm of aperiodic, nano-scale reconnection events that maintain the loop in a hot, highly dynamic regime. Outflow jets caused by magnetic reconnection resulting from braiding can also be found throughout these simulations, and some examples have already been shown in previous works (e.g., Figure~5 of \citealt{antolin2021reconnection},  Figure~8 of \citealt{de2022probing}) and in \cite{2025A&A...695A..40C}. 

Figure \ref{Fig:paper_2_scheme} sketches the sequence of events that accelerates a reconnection outflow:
photospheric motions push neighbouring flux strands one toward the other (A); there, the current density steeply grows into a thin current sheet or ``dissipation region'' (\citealt{hesse1988theoretical,schindler1988general}) where the field-line connectivity breaks and rejoins (B); finally, the newly reconnected lines are no longer in force balance, and their curved geometry stores excess magnetic tension that slingshots them outward (C). 

Figure \ref{Fig:paper_2_3D_view_1} shows three-dimensional views of the simulation domain at time $t \sim 2400$ s.  The left panel (and Movie I) shows the density stratification rising from one chromospheric footpoint, overlaid with four colour-coded bundles of field lines that trace the magnetic flux tubes subject to continuous photospheric twisting; the field lines are already tightly braided, underscoring the intricate connectivity produced by the imposed vortices.  In the centre panel (and Movie I), several narrow current sheets (blue) arise above towers of high temperature.  By this time the original 4 tubes have fragmented into many smaller strands, and the persistent reconnection events therein heat the plasma to an average of roughly $1.5$ MK.  In the right panel (and Movie I), a horizontal slice through the loop apex ($z=0$ Mm) presents the velocity module map; light-blue and magenta lines outline the magnetic geometry on either side of a typical current sheet, and a high-speed outflow expands from the region where the two bundles converge and reconnect, indicating the bidirectional jets expected from small-angle reconnection events (see the scheme in Fig. \ref{Fig:paper_2_scheme}). In Paper I we describe in detail the properties of the loop system in this post-avalanche  `steady heating phase', in terms of thermal and dynamic properties, temporal evolution, and synthetic observables.

\section{Methods}
\label{sec:methods}

We now describe the method devised for the automatic detection of reconnection outflow jets (ROAD: Reconnection Outflows Automatic Detection).

\begin{figure*}[ht!]
   \centering
\includegraphics[width=\hsize]{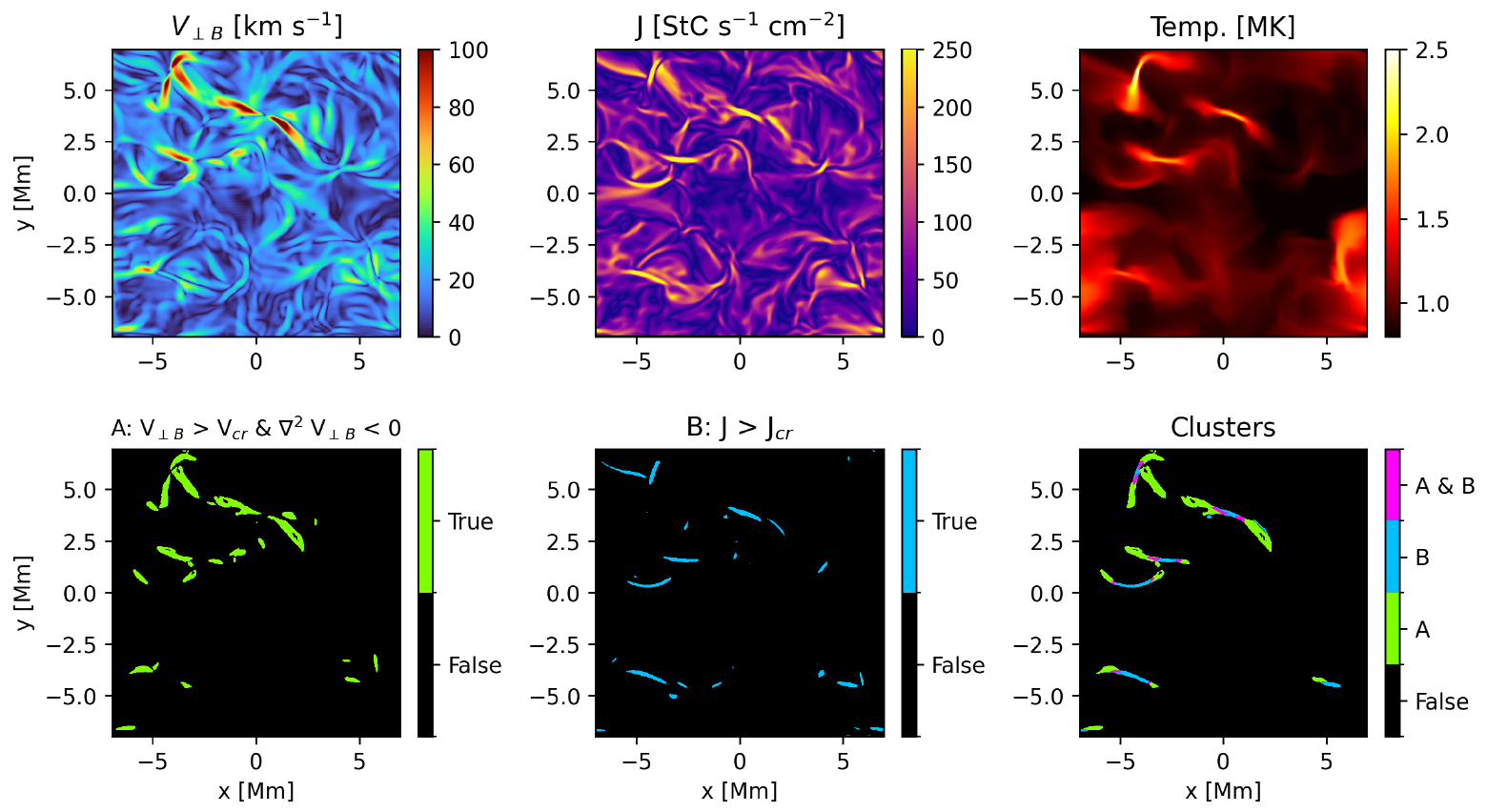}
  \caption{Example of automatic detection reconnection outflows in 3D MHD simulations. Top row: Horizontal cut of the velocity magnitude perpendicularly to the magnetic field across the midplane (corresponding to the loop apex) and at time $t \simeq 2400\,\mathrm{s}$ (left panel); cut of the current density magnitude (middle panel); and, cut of the temperature (right panel).
  Lower row: Horizontal cut of high-velocity bumps (left panel), cut of high current dissipating regions (middle panel); and,
  cut of the reconnection outflow clustering as a result of the automatic detection method (right panel). In green (blue) we show the regions where condition $\mathbb{A}$ ($\mathbb{B}$) in Eq. \ref{eq:clustering_1} (\ref{eq:clustering_2}) is fulfilled. The two regions may overlap (magenta). Movie II  shows the evolution of the six panels, including the impulsive reconnection events yielding to fast outflows (upper left), strong current sheets (upper left) and heating (upper right); and their detection by ROAD (lower panels).} 
  \label{Fig:paper_2_clustering_method}
\end{figure*}

\begin{figure}[ht!]
   \centering
\includegraphics[width=0.7\hsize]{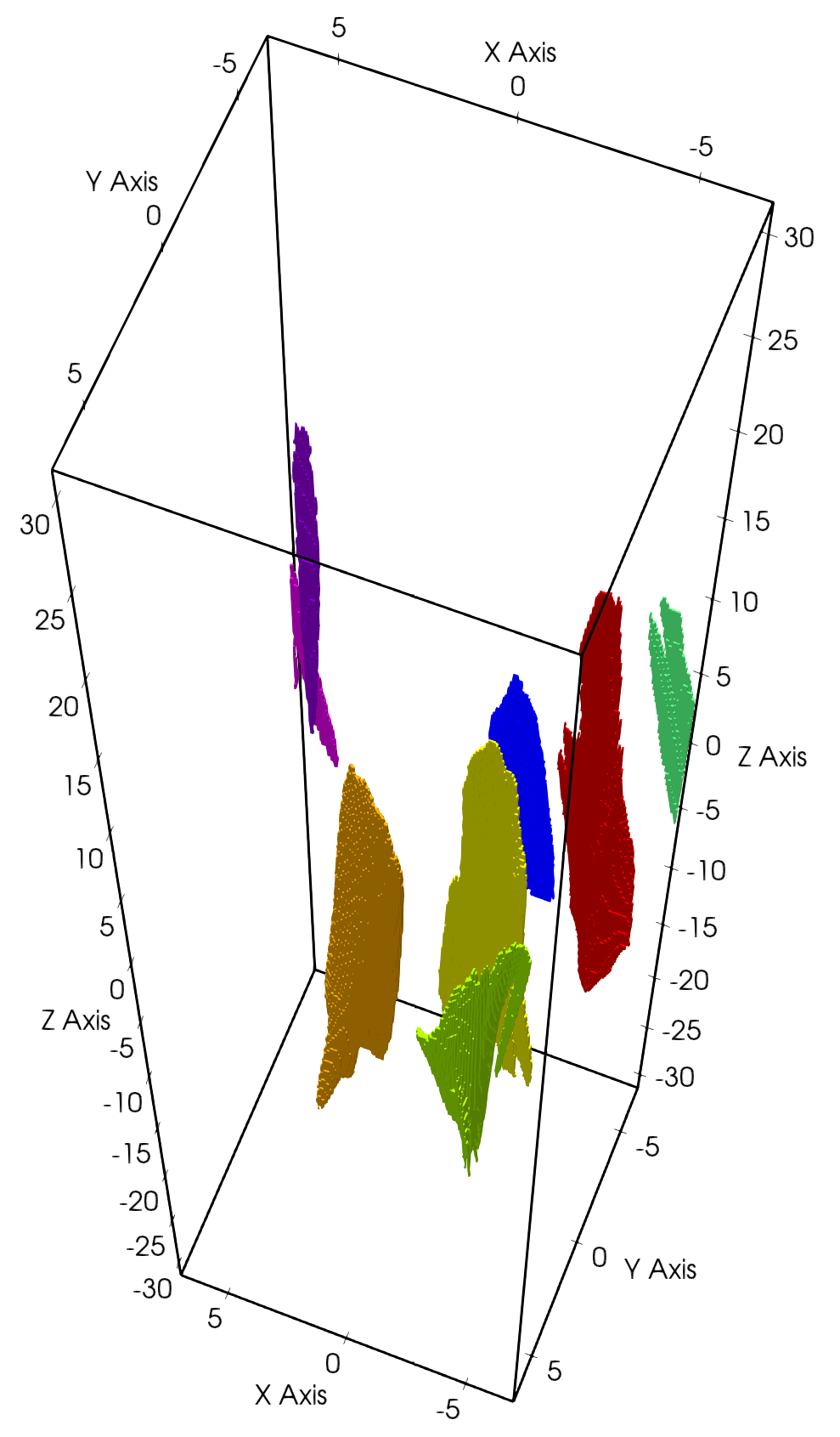}
  \caption{A 3D rendering of the clustering algorithm results $t \simeq 2400\,\mathrm{s}$. Each cluster \textcolor{black}{satisfies Eq. \ref{eq:clustering_5}} and is shown with a different colour.} 
  \label{Fig:paper_2_3D_view_2}
\end{figure}

As described in Paper I, the time-dependent, resistive MHD equations are solved for density ($\rho$), pressure ($p$), velocity and magnetic vector fields ($v_i$ and $B_i$, with $i = x\,y\,z$) with the PLUTO code \citep{mignone2007pluto},
in a Cartesian 3D box of $14\times14\times62\,$Mm$^{3}$ ($x$, $y$, $z$). 
An anomalous resistivity of magnitude $\eta_{0}=10^{11}\,\mathrm{cm}^{2}\,\mathrm{s}^{-1}$ is activated wherever the local current density exceeds $J_{\rm tr}=250\,\mathrm{StC}\,\mathrm{cm}^{-2}\,\mathrm{s}^{-1}$ only in the corona (temperature above $T_{\mathrm{cr.}} = 10^{4}\,\mathrm{K}$), to trigger impulsive reconnection in coronal thin current sheets \citep{hood2009coronal,reale20163d}.

\begin{equation}
    \eta =
    \begin{cases}
    \eta_0 & |J| \ge J_{\mathrm{tr}}  \text{ and } T \ge T_{\mathrm{cr.}} \\
    0  & |J| < J_{\mathrm{tr.}} \text{ or } T < T_{\mathrm{cr.}}
    \label{eq:anomalous_res}
    \end{cases},
\end{equation}
where the current density is given by Amp\`ere's law in terms of the curl of the magnetic field:
\begin{equation}
\vec J = \frac{c}{4 \pi} \mathbf{\nabla} \times \vec B,
\label{Eq:current_density}
\end{equation}

\textcolor{black}{The automatic detection of reconnection jets is based on specific properties of velocity and current density. 
More specifically: 1. We first extract a boolean 4D-cube (space and time) to gather the plasma volume with high values of velocity and/or current density, possibly involved in reconnection outflow events; 
2. We then apply a (connected-components) machine learning clustering algorithm \citep{he2017connected} to the new cube, to separate different events in space and time; 3. From the clustering results, we apply a selection criterion to restrict the set of events to only those where reconnection yields fast outflows.}

We define a ``reconnection outflow jet'' as a rapid plasma outflow generated by the slingshot effect of magnetic field lines after reconnection \citep{antolin2021reconnection}. 
During component reconnection within strong guide magnetic fields, plasma is likely to be accelerated by the released magnetic tension perpendicularly to the field itself.
We thus consider any region of plasma $\mathbb{A}$ (Eq.~\ref{eq:clustering_1} below) around a local maximum of the velocity field ($\nabla^2 |v_{\perp}| < 0$), and where the component of the velocity perpendicular to the magnetic field exceeds a certain threshold value (e.g., $45~\, \mathrm{km}\,\mathrm{s}^{-1}$):
\begin{equation} 
        \mathbb{A} \, : \,
    \begin{cases}
    |v_{\perp}| > 45\, \mathrm{km}\,\mathrm{s}^{-1}, \\
    \nabla^2 |v_{\perp}| < 0;
    \end{cases}
    \label{eq:clustering_1}
\end{equation}
where $v_{\perp} = \vec v - \frac{\vec v \cdot \vec B}{|B|}$ is the velocity component perpendicular to the magnetic field. As an additional condition, we consider a region of space $\mathbb{B}$ (Eq.~\ref{eq:clustering_2}) where the magnetic field is dissipated, which can be identified as the region where the current density exceeds a certain threshold \citep{hood2009coronal, reid2020determining}:
\begin{equation} 
        \mathbb{B} \, : \,
    |J| > J_{\mathrm{tr.}}. 
    \label{eq:clustering_2}
\end{equation} 
\textcolor{black}{The clustering procedure is applied to the union of the two sets, 
$\mathbb{A} \cup \mathbb{B}$, yielding a collection of connected regions 
$\{\mathbb{C}_i\}$. Among these, we retain only those clusters 
that contain contributions from both $\mathbb{A}$ and $\mathbb{B}$, 
i.e. whose intersection with each set is non-empty:
\begin{equation}
    \mathbb{C} \quad: \quad \{\,\mathbb{C}_i \cap \mathbb{A} \ne \emptyset 
    \quad \& \quad
    \mathbb{C}_i \cap \mathbb{B} \ne \emptyset \,\}.
    \label{eq:clustering_5}
\end{equation}
This criterion ensures that the selected clusters represent regions of joint 
activity where events from $\mathbb{A}$ and $\mathbb{B}$ coexist or overlap.}
Therefore, this identifies volumes of space where the plasma is accelerated as a consequence of released magnetic tension during reconnection. By clustering, we can also track the temporal evolution of each nanojet, allowing for statistical analysis of the results in terms, for instance, of: i) released magnetic/internal/kinetic energy; ii) correlations between temperature, velocity, and current density; iii) geometrical properties (such as volume and circularity) and their relationship with other physical quantities.

\textcolor{black}{In general, the velocity condition in Eq. \ref{eq:clustering_1} does not necessarily imply Eq. \ref{eq:clustering_2}, as empirically shown in the third column of Figure \ref{Fig:paper_2_velocity_treshold}, where the threshold masking is applied for both $J$ and $V_{\perp}$ at the mid plane of the box. High-speed regions generally form around current sheets but overlap only slightly with them. For this reason, it is important to consider the union of $\mathbb{A}$ and $\mathbb{B}$ before performing the clustering.}

The current threshold used throughout this study is identical to the value that activates the anomalous resistivity in the MHD equations (Eq. \ref{eq:anomalous_res}, we also discuss other methods in appendix \ref{sec:appendix2}). \textcolor{black}{The top panels of figure \ref{Fig:paper_2_velocity_treshold} illustrate the thinning of the current sheets once the current density approaches and exceeds the given threshold. In particular, in the cumulative volume fraction of plasma with a current density below a given value (left plot), the curve breaks into two distinct slopes at  $J \sim J_{tr} = 250\,\mathrm{StC}\,\mathrm{s}^{-1}\,\mathrm{cm}^{-2}$.}

The velocity threshold is, instead, derived empirically from the statistics of the flow field.  The lower (left) panel of figure \ref{Fig:paper_2_velocity_treshold} plots the cumulative volume fraction of plasma with speeds below a given value.  The curve breaks again into two distinct slopes at $V_{\mathrm{tr}} \simeq 45\, \mathrm{km\,s^{-1}}$.  For $V < V_{\mathrm{tr}}$ the enclosed volume decreases steeply for increasing $V$, reflecting the broad, turbulent motions generated mainly by the persistent footpoint stirring.  Above $V_{\mathrm{tr}}$ the slope flattens significantly.
The right panels of Figure \ref{Fig:paper_2_velocity_treshold} clearly show that the change of the slope is associated with the local acceleration of the outflows by magnetic reconnection.  We therefore adopt $V_{\mathrm{tr}}$ as a self-consistent boundary that separates reconnection-driven jets from the ever-present turbulent background.  Because occasional turbulent eddies can still produce isolated high-speed spikes, we further require that candidate volume cells satisfy the current threshold criterion as well; only the compact regions that meet both conditions (Eqs. \ref{eq:clustering_1} and \ref{eq:clustering_2}) are classified as ``true'' reconnection outflows (Eq. \ref{eq:clustering_5}).
\textcolor{black}{Moreover, while the adopted velocity threshold is sufficiently high to exclude most turbulence-driven perturbations, it is still low enough to capture the full 
evolution of reconnection outflows, from their initial impulsive phase to the later 
stages of decay. As shown in Figure \ref{Fig:paper_2_velocity_treshold} (leftmost panel), adopting higher values of $V_{\perp}$ would remove outflow events whose peak velocities are modest or whose formation occurs farther from the main current sheet. According to the algorithm’s criteria, the relatively slow regions must remain connected to dissipating current sheets, ensuring that reconnection is still active, albeit with a progressively decreasing rate. On the other hand, when extracting statistical results, it is crucial to 
consider the temporal evolution of each cluster to distinguish genuinely slow 
outflows from mature reconnection regions that are gradually dissipating and 
therefore decelerating.}

Figure \ref{Fig:paper_2_clustering_method} and the accompanying Movie II present a horizontal slice through the box mid-plane, i.e., the loop apex, (z = 0 Mm) at $t \sim 2400$ s.  The upper row  shows that bidirectional, high-speed outflows, evident as paired red streaks in the velocity map, are frequently co-located with intense, sheet-like currents and high temperatures, suggesting ongoing small-angle reconnection.  The lower row shows how these streaks are pinpointed, from the left, pixels where: a) the flow speed exceeds the high-velocity threshold (Eq. \ref{eq:clustering_2}), b)  the current density exceeds the threshold (Eq. \ref{eq:clustering_1}), c)  both conditions are met simultaneously.  The resulting clusters typically consist of a compact core of strong current surrounded by one or more “wings’’ of rapidly expanding plasma.  

Figure \ref{Fig:paper_2_3D_view_2} provides a fully three-dimensional view of the feature-clustering result at $t \sim 2400$ s.  Each cluster \textcolor{black}{satisfies Eq. \ref{eq:clustering_5}} and is rendered in a distinct colour, making their individual morphologies easy to distinguish within the box.  The objects appear as slender, sheet-like volumes that extend preferentially along the background (guide) field direction, while remaining only a few grid cells thick in the transverse plane.  This geometry underscores the picture of reconnection taking place in narrow, elongated current sheets braided through the loop apex.

\section{Results}
\label{sec:results}

\begin{figure}[pt]
   \centering
\includegraphics[width=0.9\hsize]{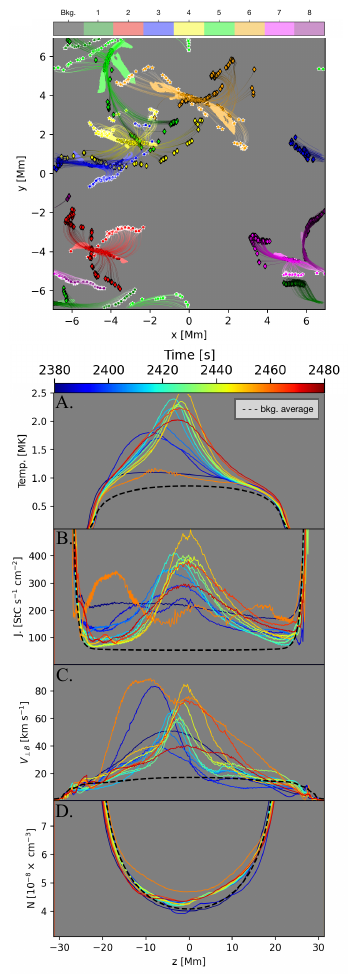}
  \caption{Reconnection outflow properties along the guide field. Top map: horizontal cut of the 3D box shown in Fig. \ref{Fig:paper_2_3D_view_2}. The detected reconnection outflows are identified by the eight colour-coded clusters. Field lines crossing each cluster are projected onto the horizontal plane and the upper (lower) footpoints pinned by white-contoured stars (black-contoured diamonds). Lower panels: from the top to bottom, field lines averaged values of temperature, current density, velocity and plasma density along the loop length ($z = 0$ is the apex) for cluster n. 5. See Appendix for more results.} 
  \label{Fig:paper_2_composition}
\end{figure}

Figure \ref{Fig:paper_2_composition} follows the temporal evolution of individual reconnection outflows in greater detail. Here we discuss a representative example (cluster 5) that evolves roughly between $2400$ and $2450\,\mathrm{s}$, while Appendix A1 presents analogous results for a few additional clusters within a similar time range, which exhibit the same overall behaviour.  The top panel displays a horizontal cut at the loop apex ($z = 0$ Mm) at $t \sim 2400$ s, with each coloured patch marking a different outflow cluster identified by the selection criteria of Eq. \ref{eq:clustering_5}.  \textcolor{black}{Inside the dissipation core (or current sheet, region $\mathbb{B} \cap \mathbb{C}_i$) of every event, we seed, at every snapshot time, 64 points at random positions. Field lines are then traced from the midplane using a second-order Runge-Kutta integration scheme, and mapped into the lower and upper sides of the box;} their projections onto the horizontal plane are over-plotted, with diamonds and stars indicating the upper and lower footpoints. The strong, localised gradients in field line connectivity are characteristic of three-dimensional reconnection sites. \textcolor{black}{In fact, here, we examine the magnetically connected regions where field lines are reconnecting, specifically where impulsive heating occurs and plasma is rapidly accelerated.}

Along each traced line we sample temperature (panel A), current density (panel B), flow speed (panel C), and plasma density (panel D), and then average these quantities as functions of distance along the loop axis.  The resulting profiles, shown in the lower panels as a function of time, reveal that the temperature and current density peak sharply at nearly the same location, defining the compact core of the outflow.  Both quantities rise rapidly at the onset of the event and subsequently decrease toward their background values \textcolor{black}{(dashed black curves, obtained by averaging outside of the reconnection outflows regions)}.  The velocity profile is likewise initially confined near the apex but widens over several tens Mm as the outflow propagates along the guide field.  Density changes are more gradual: a modest increase appears during the late stages of the energy release, consistent with chromospheric evaporation filling the loop.

\begin{figure*}[h!]
   \centering
\includegraphics[width=\hsize]{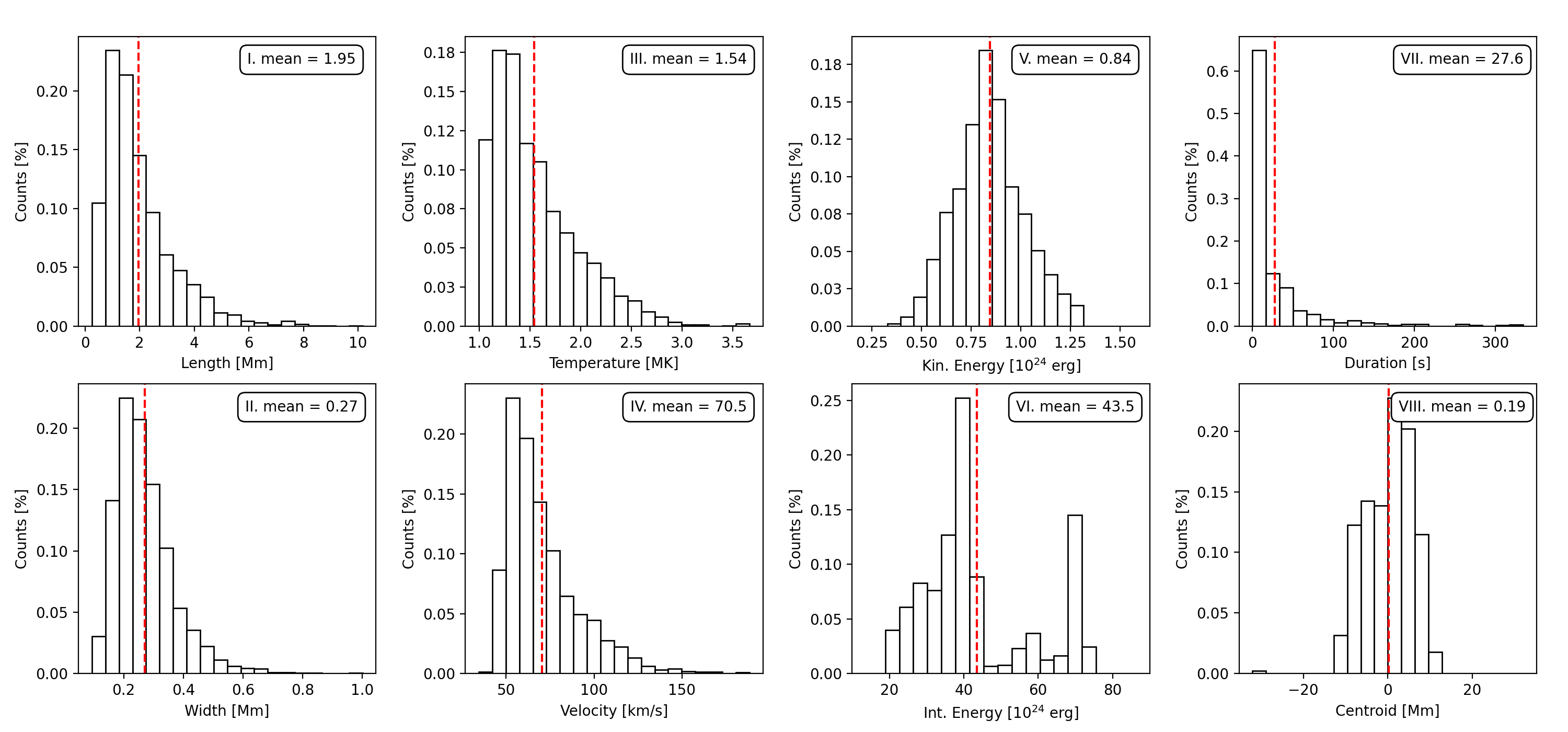}
  \caption{Statistics of reconnection outflows properties derived from the simulation. From top left to lower right, respectively: histograms of the length, temperature, kinetic energy, duration, width, velocity, internal energy, and centroid distributions of the detected reconnection outflows. Average values are marked (red dashed lines) .} 
  \label{Fig:paper_2_statistics_2}
\end{figure*}

Figure~\ref{Fig:paper_2_statistics_2} summarizes the statistical properties of all the reconnection outflows detected during the simulation ($\sim 4000\,\mathrm{s}$ long). The eight panels of histograms display: length (I) and width (II) of the clusters $\mathbb{C}$, measured from the horizontal cut at their centroid height (see below); the peak plasma temperature (III) and maximum flow speed (IV); the kinetic (V) and internal (VI) total energies released; and the event duration (VII) together with the centroid position (VIII) along the loop axis ($\hat z$).

The outflow dimensions are obtained from a horizontal slice through each cluster at the height of its geometric centre.  Approximating the slice as a rectangle of area $A$ and perimeter $P$,  the ``length'' ($L$) and ``width'' ($W$) of the rectangular area can be expressed as:
\begin{equation}
    L = \frac{P}{4} \left[1+\sqrt{1-\frac{16A}{P^{2}}}\right],\quad   
    W = \frac{P}{4} \left[1-\sqrt{1-\frac{16A}{P^{2}}}\right],
\end{equation}
with $L>W$.  
The resulting distribution shows that the structures are extremely slender, with a typical aspect ratio $W/L \sim 0.1$.

Physically, the jets are hot and fast: temperatures exceed $1.5$ MK and reach up to $\sim3$ MK, while flow speeds range from $45$ to $150$ km s$^{-1}$ with a mean of $\sim70$ km s$^{-1}$.  Their energy content is firmly in the nanoflare regime; the internal‐energy release outweighs the kinetic contribution by roughly two orders of magnitude.  Most events persist for only a few tens of seconds, although a handful last longer than two minutes.  Finally, the centroid histogram shows that activity is concentrated near the loop apex: the distribution peaks sharply at $z=0$ Mm and rarely extends beyond $\pm15$ Mm along the axis. 

Magnetic braiding, and the currents it generates, tend to be the strongest well above the loop footpoints. In long, thin, low‑$\beta$ coronal flux tubes, the large Alfvén speed efficiently mitigates asymmetries along the guide field direction 
\citep{cozzo2023asymmetric}, while the large aspect ratio allows for small-angle field-line misalignments only. By contrast, plasma near the chromosphere is denser and the magnetic field is rooted in a high‑$beta$ environment (making heating and outflow acceleration harder, also considering that field lines have little room to expand). Moreover, the strong magnetic tension prevents the formation of narrow current sheets \citep{rappazzo2013current}.
Altogether, these factors suggest the region around the apex to be the most likely site for efficient component reconnection and associated high‑speed outflows.

\begin{figure*}[h!]
   \centering
\includegraphics[width=\hsize]{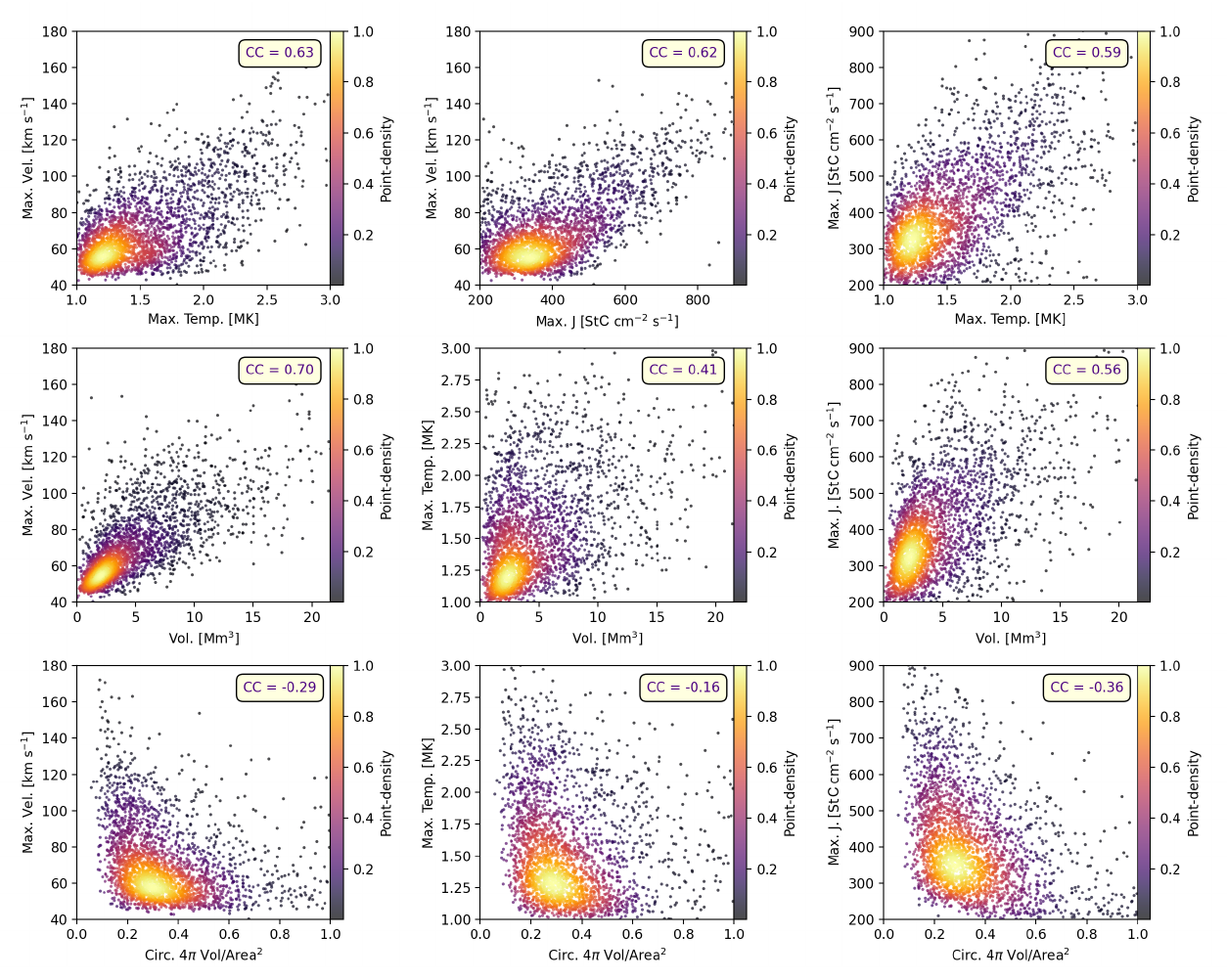}
  \caption{2D histograms of outflows physical and geometric properties, highlighting possible correlations between variables. 
  Top row (from left): Outflows maximum velocity v.s.\ maximum temperature; maximum velocity v.s.\ maximum current density; and maximum current density v.s.\ max temperature. 
  Middle row (from left): Outflows maximum velocity, v.s.\ volume; nanojets maximum current density, v.s.\ volume; nanojets maximum temperature, v.s.\ volume.
  Bottom row: Outflows maximum velocity v.s.\ circularity; and maximum current density v.s.\ circularity, maximum current density v.s.\ temperature.} 
  \label{Fig:paper_2_statistics_1}
\end{figure*}

Figure \ref{Fig:paper_2_statistics_1} quantifies how the physical properties of the identified reconnection outflows relate with each other and to their geometry.  For all the automatically detected reconnection outflows, we plot their population density distribution of each event $\mathbb{C}$, for two variables in each plot, and explore the correlation between these variables. In particular, we consider:  maximum velocities, temperature, and current density, as well as volume \textcolor{black}{of each clustered region} and shape factor, or ``circularity'', defined as
\begin{equation}
 \mathrm{Circ} = \frac{12\pi\,\mathrm{Volume}^2}{\mathrm{Area}^3},   
\end{equation}
so that $\mathrm{Circ.} = 1$ for perfect spheres and $\mathrm{Circ.} \ll 1$ for elongated structures.
A clear, positive trend links outflow temperature, current density and speed: larger current densities produce hotter clusters and faster plasma motions, and all three quantities increase systematically with the cluster volume.  \textcolor{black}{As previously shown in Figure \ref{Fig:paper_2_velocity_treshold}, the spatial extension of the reconnection outflows is also dependent of the chosen values of current and velocity thresholds, with lower levels yielding to bigger clusters.} In contrast, circularity anticorrelates with these same parameters: the hottest and fastest outflows (those anchored in the strongest current sheets) appear more elongated and sheet-like, yielding smaller values of $\mathrm{Circ}$, confirming their jet-like character.

\begin{figure*}[h!]
   \centering
\includegraphics[width=\hsize]{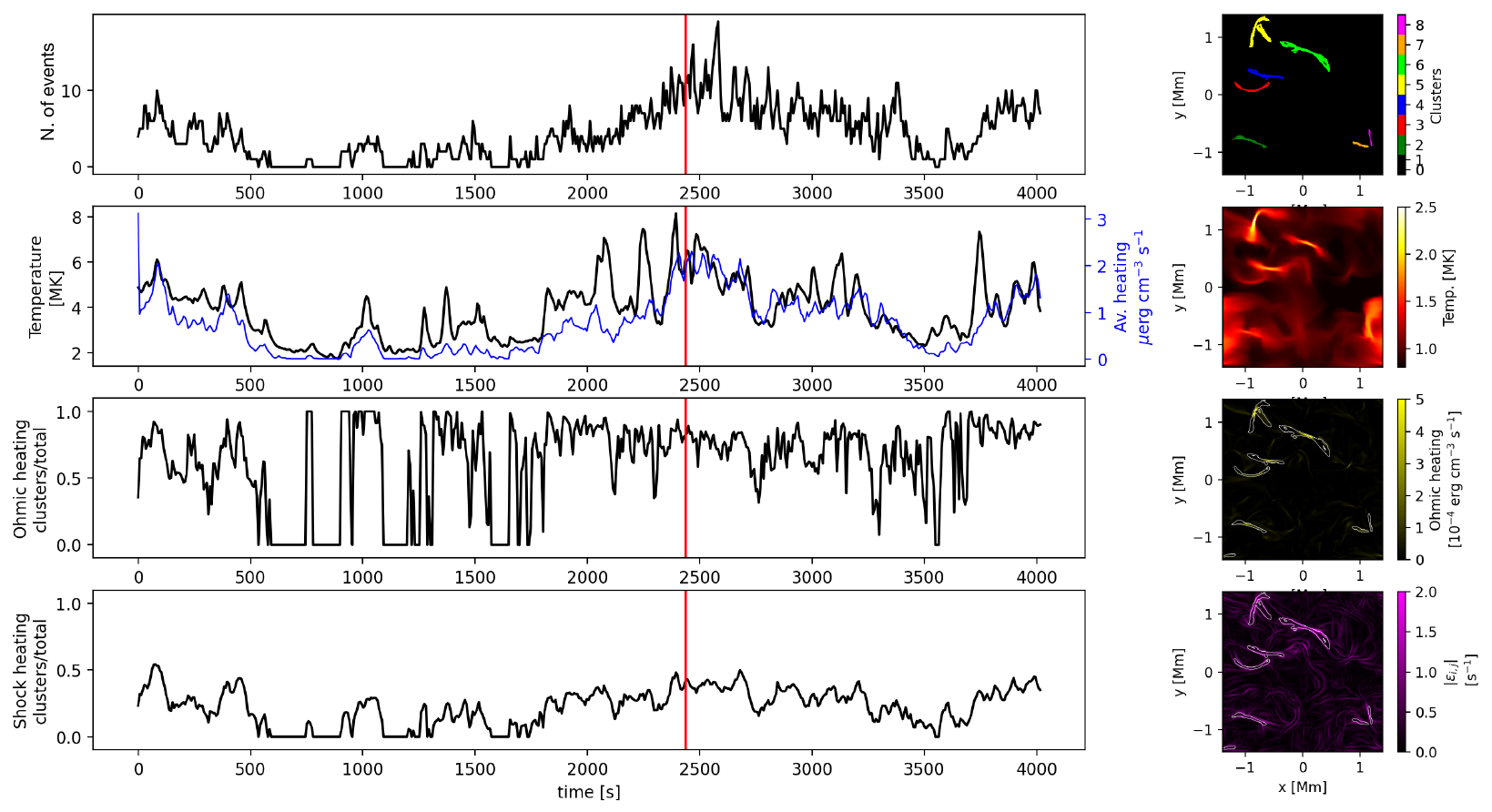}
  \caption{Occurrence of reconnection outflows through the time and their impact on the coronal heating. First row from the top panel: Temporal evolution of the instantaneous number of detected events (left), horizontal cut of the clustering (right) at $t \simeq 2400\,\mathrm{s}$ (red line on the left). Second panel: Temporal evolution of maximum temperature and averaged volumetric heating (black and blue curves on the left plot), horizontal cut of the temperature (right). Third row: ratio of clustered v.s. total instantaneous ohmic heating as function of time (left), horizontal cut of the ohmic heating distribution (right). Fourth row: estimated clustered v.s. total instantaneous shock heating (left) as function of time (left),  horizontal cut of the strain hydrodynamic tensor modulus (right).} 
  \label{Fig:paper_2_time_evolution}
\end{figure*}

Figure \ref{Fig:paper_2_time_evolution} links the statistics of the reconnection outflows to the global energetics of the model corona. The upper-left panel tracks the number of identified events in each snapshot; their occurrence rate is highly intermittent, mirroring the bursty nature of the driver.  
The temporal profiles of the maximum temperature and the total Ohmic heating rate 
both rise and fall in concert, demonstrating that the hottest episodes coincide with periods of enhanced magnetic dissipation.

The two lower panels on the left compare, for every snapshot, the Ohmic and shock-viscous heating produced ``within'' the clustered volumes against the corresponding domain-integrated values.  Although the MHD equations solved by PLUTO 
do not explicitly include viscous heating in shocks, we consider it, a posteriori, to be proportional to the norm of the hydrodynamic stress tensor $\epsilon$,
\begin{equation}
|\boldsymbol{\epsilon}| = \left(\sum_{i,j}\epsilon_{ij}^{2}\right)^{1/2},
\qquad
\epsilon_{ij} \ \propto \ \partial_{i}v_{j}+\partial_{j}v_{i}-\tfrac23\delta_{ij}\nabla\!\cdot\!\mathbf{v}.
\end{equation}
The clustered-to-total shock heating ratio shown in the lower panel does not depend on the specific value of the numerical viscosity, which is unknown.
The maps (on the right) are a horizontal cut through the loop apex that shows the cluster mask, temperature, Ohmic-heating density, and $|\boldsymbol{\epsilon}|$ at $t \sim 2400$ s, and provide a visual reference.  Nearly all strong Ohmic sources fall within the selected jets, whereas more than half of the instantaneous shock heating originates outside of those volumes. The Ohmic and shock heating curves in  Fig. \ref{Fig:paper_2_time_evolution} show similar trends, with peaks that correlate well with the number of detected events. As expected, reconnection outflows impact both ohmic and viscous heating, although with different relative weights. In particular, Ohmic contribution to coronal heating is dominated by dissipating current sheets leading to reconnection outflow jets. On the other hand, heating by shock-viscous dissipation arises partly also from unrelated turbulent motions and waves evenly distributed throughout the corona that add to the localized reconnection outflows. 
\textcolor{black}{It is important to note that viscosity is not explicitly included in our MHD model. As a consequence, the quantitative estimation of shock heating is subject to some uncertainty. The effective shock viscosity in the adopted numerical schemes is inherently resolution-dependent. The derived shock-heating ratio should therefore be interpreted as an approximate diagnostic. A more accurate assessment would require the implementation of an explicit viscosity which is beyond the scope of the present work but will be explored in future studies.} For example, higher Prandtl numbers, arising from an explicit treatment of the shock viscosity, would eventually smooth high-speed flows and thicken current sheets \citep{park1984reconnection}.

\section{Discussion and Conclusions}
\label{sec:conclusions}
In this work, we described and tested ROAD, an algorithm for the automatic detection of component-reconnection events that yield fast outflows.

With this algorithm, we investigated coronal heating through magnetic braiding of a coronal loop, modeled using numerical simulations of fully compressible plasma from three-dimensional MHD equations (\citealt{mignone2007pluto}). Twisting motions at the loop's footpoints energize the magnetic field, leading to turbulent non-linear dynamics and the continuous formation of field-aligned current sheets. As a result, the corona is heated through brief, localized bursts, and rapid reconnection drives outflows. Current sheets dissipate  unevenly, so that only a fraction of the coronal mass and volume is heated at a time (see \ref{Fig:paper_2_3D_view_1}). Indeed, both temperature and current density show fine-scale, sub-arcsec structures, close to or even below current observational capabilities of solar EUV instruments that observe plasma at $1$-$3\,\mathrm{MK}$.

To detect candidates of reconnection jets in the simulated solar atmosphere, we base our criteria on physical assumptions, requiring, together with the acceleration of strong outflows perpendicular to $\vec B$ (Eq. \ref{eq:clustering_1}), also the presence of reconnecting current sheets, identified by regions with strong current density (Eq. \ref{eq:clustering_2}).
The detected events, although different in size and morphology, broadly show a sheet-like structure (Fig. \ref{Fig:paper_2_3D_view_2}), with a narrow cross-section consisting of a central current sheet, enveloped by rapid-flow wings (Fig. \ref{Fig:paper_2_clustering_method}).
Along the field-aligned direction, the detected outflows show narrow temperature, current density, and velocity distribution, peaked around the loop apex (Fig. \ref{Fig:paper_2_composition}, and \ref{Fig:paper_2_temp_lines} to \ref{Fig:paper_2_dens_lines}, in appendix \ref{sec:appendix2}). They rapidly depart from localized current sheets, distributed mainly around the loop apex (Fig. \ref{Fig:paper_2_statistics_2}, \citealt{antolin2021reconnection}), that dissipate and eventually make coronal plasma density increase by chromospheric evaporation. 

\cite{sukarmadji2022observations} and \cite{gao2025reconnection} show the statistical results from observations of several reconnection nanojets. In particular, \cite{sukarmadji2022observations} report a number of IRIS and SDO/AIA observations of nanojets observed in different coronal loop structures, including one powered by a blowout jet, and two associated with coronal rain, while \cite{gao2025reconnection} catalogue 27 nanojets from an untwisting filament observed with Solar Orbiter/EUI’s on 30 September 2024 with $2\,s$ cadence and sub-arcsecond resolution. The typical velocities are inferred by \cite{sukarmadji2022observations} around $150\,\mathrm{km}\,\mathrm{s}$, while the jets observed by \cite{gao2025reconnection} are markedly faster than previously reported, most of them exceeding $450\,\mathrm{km}\,\mathrm{s}^{-1}$. However, the two ensembles of events show similar kinetic ($\sim 10^{24}\,\mathrm{erg}$) energies. Statistical results are also comparable in terms of duration of the events ($\sim 20\,\mathrm{s}$) and dimensions (widths $\sim 0.6 \,\mathrm{Mm}$ and lengths $\sim 2\,\mathrm{Mm}$, as reported by \citealt{sukarmadji2022observations}).
The average values and distributions inferred from observations broadly mirror the statistical results shown in Fig. \ref{Fig:paper_2_statistics_2}. However, we note that a direct, one-to-one comparison of the observed nanojet properties with the properties of reconnection outflows we derive from our simulation is not applicable and it is beyond the scope of this paper: in this work, we extract intrinsic physical quantities from the simulations, which cannot be directly compared with observed quantities. In a follow-up work, we will analyze synthetic observables from the simulation, and devise a corresponding automatic detection algorithm to derive outflow properties that can be more directly compared with observations.  

The detected events display the key signatures expected from reconnection theory, as also shown in Fig. \ref{Fig:paper_2_hs}, discussed in appendix \ref{sec:appendix2}, and closely resemble the small-angle reconnection scenario of \citet{antolin2021reconnection}, also further addressed by \cite{2025A&A...695A..40C}. Both cases produce short-lived ($\sim 30\,\mathrm{s}$) reconnection outflow jets of Mm size and speeds of a few hundred $\mathrm{km}\,\mathrm{s}^{-1}$, expanding from narrow current sheets between two, slightly tilted bundles of field lines. While observations generally reveal uni-directional jets \citep[e.g.,][]{patel2022hi}, the detected outflows generally expand in opposite directions, although their irregular shape often shows evident asymmetries. Since the simulation does not account for the intrinsic magnetic curvature of coronal loops \citep{pagano2021modelling}, only local uniformities of the surrounding magnetic field and plasma can distort the propagation of the outflows on the two sides in different ways.

The presented results also broadly agree with the 2.5-D MHD simulation of a current sheet within a guide-field described by \cite{sen2025merging}. They indeed examine how impulsive velocity perturbations can fragment a coronal current sheet into multiple plasmoids that later merge and produce, at their interface, short-lived ($\sim\,10\mathrm{s}$), highly collimated ($500\,\mathrm{km}$) bidirectional jets that reach $\sim 100\,\mathrm{km}\,\mathrm{s}^{-1}$ and carry
$\sim{3 \times 10^{24}\,\mathrm{erg}}$. The resulting flows share with the reconnection outflows presented here similar key kinematic, morphological, and energetic signatures. 

Field lines originating from the selected current sheets show, in general, strong connectivity gradients, corresponding to high ($Q \gg 1$) squashing factors (Fig. \ref{Fig:paper_2_composition}, and \ref{Fig:Squashing_factor} in appendix \ref{sec:appendix2}). Moreover, they form coherent structures, not  dissimilarly from the current sheet connected (CSC) regions discussed by \cite{rappazzo2017coronal}.

The ohmic heating captured by the outflows covers most of the heat released by anomalous diffusivity (Fig:\ref{Fig:paper_2_time_evolution}), suggesting nanojets, and in general reconnection jets, play an important role in impulsive heating processes. Vice versa, ohmic heating can be considered a fair proxy for reconnection processes, as already outlined by \cite{reid2020determining}.

\textcolor{black}{
The geometrical aspect ratio of the reconnecting regions 
is 
comparable to the ratio between inflow and outflow velocities,  
as expected from mass conservation, assuming negligible plasma 
compressibility. 
For comparison, we considered a steady-state Sweet--Parker configuration with 
$B_{\mathrm{rec}} = 2~\mathrm{G}$, 
$\rho = 2 \times 10^{-15}~\mathrm{g}\,\mathrm{cm^{-3}}$, 
$\eta = 10^{11}\,\mathrm{cm^{2}\,s^{-1}}$, and 
$L = 1\,\mathrm{Mm}$. The predicted Alfvén speed 
($V_A \approx 140\,\mathrm{km\,s^{-1}}$) is comparable to the measured outflow velocities, inflow velocity and current sheet width
($V_{\mathrm{in}} \approx 0.3\,\mathrm{km/s}$,
$\delta \approx 2\,\mathrm{km}$) are considerably smaller. 
These differences are expected, as the simulated reconnection events are transient and occur in a component-reconnection full-3D regime. The estimated Lundquist number 
($S = 4\pi L V_A / \eta \approx 2\times10^{5}$) 
places the system close to the threshold for plasmoid instability. 
However, the characteristic reconnection timescales are shorter than those required 
for secondary tearing to develop \citep{sen2025merging}.
Previous comparative studies 
\citep{faerder2023comparative, faerder2024comparative} 
have shown that, while different resistivity prescriptions (including anomalous 
diffusion) can alter the small-scale morphology of the current sheet, the large-scale 
thermodynamic evolution 
remains largely unaffected. 
Here, we use anomalous magnetic diffusivity to mimic the impulsive, 
localized nature of coronal reconnection within with full-3D
numerical resolutions. Since our analysis focuses on the statistical properties 
of the resulting reconnection outflows rather than on the microphysics of current 
sheets, the main results are not expected to be sensitive to the missing small-scale 
physics.
}

One limitation of this work is the ad-hoc bottom driver and it is of crucial interest to bring this analysis to other 3D Radiative MHD models and for various solar targets.
In particular, in the simulation, we consider a long-lasting but slow photospheric driver (twisting) that is superimposed on higher-frequency but weaker random perturbations (see Paper I).
The spatial distribution of the detected heating events does not show any correlation with the adopted driver. On the other hand, the frequency and intensity of the energy release are expected to be influenced by the driver.
Impulsive DC events are released when longer driving time-scales are taken into account. Those DC heating events are easier to identify and detect in simulations as they tend to be larger, more long-lived, and less frequent than AC events. Moreover, they are expected to have a higher impact on the overall coronal loop heating, as longer driving time-scales lead to a greater time-averaged Poynting flux  \citep{howson2022effects}. 

The present paper has focused on the methodology: we introduced an automated technique that identifies reconnection-driven outflows in 3-D MHD data and we quantified their intrinsic plasma properties (sizes, lifetimes, energies, and kinematics) without yet trying to directly compare them with the nanojets reported by IRIS, AIA, or Solar Orbiter observations. 
Follow-up papers will address several open questions naturally arising from this work, to ultimately constrain what kind of environments can likely produce readily detectable nanojets, and address the question of why these nanojets are apparently not 
easily observed in images of the corona from current instrumentation.

In particular, we will address observability by forward-modelling the same simulation with chromospheric radiative transfer and instrumental responses to create synthetic IRIS, AIA, and EUI data, and testing whether the synthetic observables exhibit features comparable to the observed nanojets. Additional work in progress is aimed at applying the automatic detection algorithm to 3D radiative MHD simulations of the solar atmosphere such as Bifrost \citep{Gudiksen2011}, MURaM \citep{rempel2016extension}, and LaRE3D \citep{arber2001staggered} models. This will also provide a cross-code sanity check to explore how different numerical treatments, physical processes included,  and boundary conditions affect jet statistics. Finally, a systematic parameter survey (field strength, driving cadence, resistivity prescriptions, etc.) will clarify how nanojet occurrence and energetics scale with the underlying heating mechanism. 

Although conclusive answers will require a broader study, the present work lays the groundwork for pinpointing the diagnostic potential of reconnection outflows for coronal heating processes.

\begin{acknowledgments}
      GC and PT were supported by contract 4105785828 (MUSE) to the Smithsonian Astrophysical Observatory, and by NASA grant 80NSSC21K1684.
      PP, and FR acknowledge support from ASI/INAF agreement n. 2022-29-HH.0. 
      This work made use of the HPC system MEUSA, part of the Sistema Computazionale per l'Astrofisica Numerica (SCAN) of INAF-Osservatorio Astronomico di Palermo.
      The simulations have been run on the Pleiades cluster through the computing projects s1061, s2601, from the High-End Computing (HEC) division of NASA.
      We acknowledge the CINECA award under the ISCRA initiative, for the availability of high performance computing resources and support.
      BDP and JMS were supported by NASA contract 80GSFC21C0011 (MUSE) and NNG09FA40C (IRIS).
\end{acknowledgments}

\software{PLUTO \citep{mignone2007pluto}}

\appendix

\section{Comparative approaches to current sheets detection in MHD simulations}
\label{sec:appendix2}
This paper introduces a broadly applicable procedure for identifying reconnection outflows in 3D MHD simulations. In Sec. \ref{sec:methods} and \ref{sec:results} we described and validated the pipeline using PLUTO runs where magnetic dissipation is treated through anomalous magnetic resistivity. Because other codes may handle resistivity differently (whether through numerical diffusion or explicit resistive terms in the MHD equations) here we outline how the same recipe can be adapted, with only minor tweaks, to simulations that employ alternative resistivity treatments.
In particular, here we detail alternative methods to select the $\mathbb{A}$ zones (see, e.g., Fig. \ref{Fig:paper_2_clustering_method}) within 3D magnetic configurations, namely the sites where current sheets are expected to dissipate and ultimately trigger magnetic reconnection.

Unlike in 2-D configurations, where reconnection is confined to easily identified X-type nulls, the sites of reconnection in a fully three-dimensional magnetic field are difficult to constrain. 
Many attempts to forecast where current sheets, and therefore reconnection, will arise have relied on the connectivity gradients of magnetic field lines, quantified by following the end points of neighbouring lines \citep[e.g.,][]{priest1995three,demoulin1996three,aulanier2005current}.  

\begin{figure}[ht!]
   \centering
\includegraphics[width=0.8\hsize]{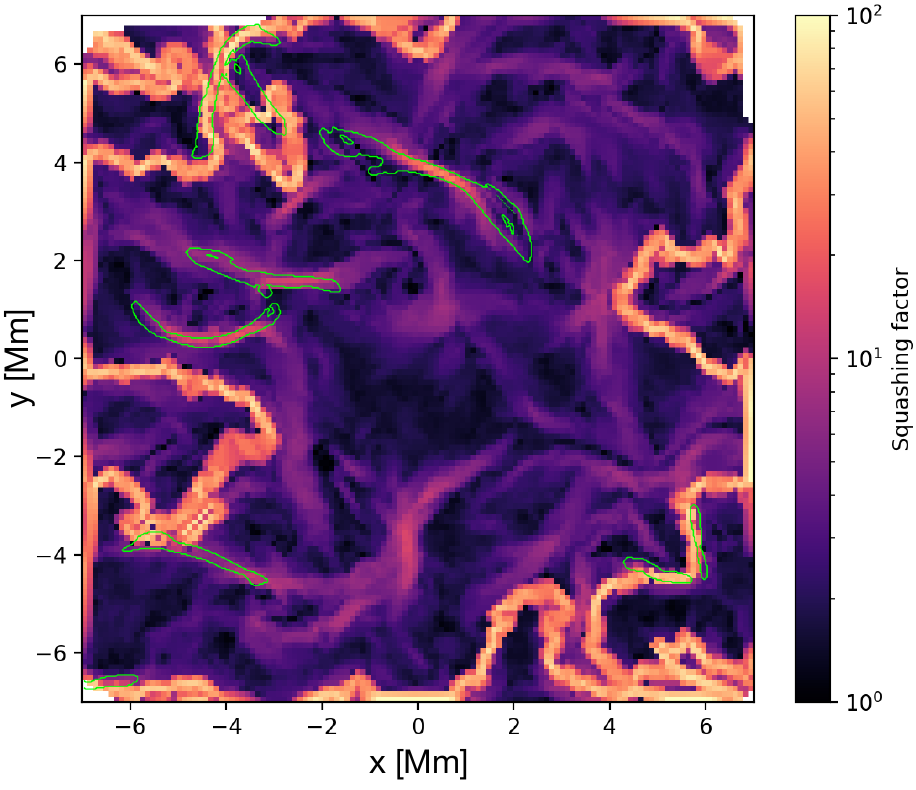}
  \caption{Squashing factor computed at the mid-plane ($z=0\,\mathrm{Mm}$).} 
  \label{Fig:Squashing_factor}
\end{figure}

A widely adopted metric is the quasi-squashing factor $Q$ \citep{titov2002theory}, which provides a direction-independent measure for the spatial gradient of the magnetic field footpoint mapping. 
Although high-$Q$ layers correlate well with reconnection sites in configurations that contain null points, their predictive power weakens in strongly braided fields with a dominant guide component.
\cite{reid2020determining} showed that in such cases $Q$ correlates less reliably with topological change than do energetics-based diagnostics such as localized Ohmic heating (see also \citealt{galsgaard2003magnetic} and \citealt{de2006numerical1,de2006numerical2}).
We draw the same conclusions, as suggested by Fig. \ref{Fig:Squashing_factor}  where the squashing factor computed at the mid-plane ($z=0\,\mathrm{Mm})$ is shown. The green contours, marking the positions of the reconnection outflows, weakly correlate with regions of high $Q$.

In general, in fully-3D reconnection processes (namely when the magnetic field does not vanish) the magnetic field reconnects inside a dissipation region where $\vec E \cdot \vec B \ne 0$ \citep{hesse1988theoretical, schindler1988general}.
In particular, \cite{hesse1988theoretical} showed that a necessary and sufficient condition for finite-B, global reconnection (i.e., where the breakdown of magnetic connection occurs for plasma elements that lie beyond the dissipation region) is:
\begin{equation}
    U_{\mathrm{fl}} = \int_{\mathrm{fl}} E_{\parallel} ds \ne 0
    \label{Eq:U_schindler}
\end{equation}
with $s$ denoting the arc length of the field line (fl), on a measurable set of field lines.
The electric field $\vec E$ here is defined by Ohm's law: 
\begin{equation}
\vec E = -  \frac{\vec v}{c} \times \vec B + \frac{\vec J}{\sigma},
\end{equation}
where $\sigma = \frac{c^2}{4 \pi \eta}$ is the electrical conductivity, $\eta$ the magnetic resistivity, and $c$ the speed of light.

\begin{figure}[ht!]
   \centering
\includegraphics[width=0.8\hsize]{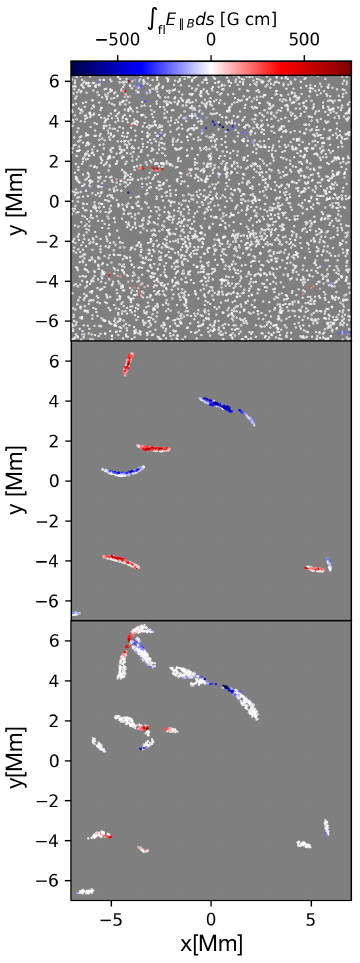}
  \caption{Distribution of electric field line-integral. We estimated the line integral of the electric vector field ($U_{\mathrm{fl}}$, Eq. \ref{Eq:U_schindler}) along 4096 field lines, with random points. In each panel, we show the point-locations where the field lines intersect the mid-plane (gray background). Each dot is coloured according to the relative $U_{\mathrm{fl}}$-value. The scattered points are grouped by whether their associated field lines pass through reconnection‑outflow current sheets (middle panel), traverse high‑velocity flows (lower panel), or lie elsewhere (upper panel).}
  \label{Fig:paper_2_hs}
\end{figure}

Fig. \ref{Fig:paper_2_hs} shows the distribution of $U_{\mathrm{fl}}$ along the mid-plane of the simulation box at time $t \sim 2400\,\mathrm{s}$. In three panels, from top to bottom, values of the electric field line integral are computed for each panel by tracing field lines passing through reconnection-outflows-associated current sheets (\textcolor{black}{top panel;} as in Eq. \ref{eq:clustering_2}), high-speed flows (\textcolor{black}{mid panel}; as in Eq. \ref{eq:clustering_1}), and elsewhere \textcolor{black}{(lower panel)}, respectively. \textcolor{black}{The field lines are traced using a second-order Runge-Kutta integration scheme}. According to the condition in Eq. \ref{Eq:U_schindler}, reconnection clearly occurs, as expected, at the current sheets. Residual reconnection also involves the inner part of the outflows' tail, as shown in the middle panel of Fig. \ref{Fig:paper_2_hs}. Anywhere else, $U_{\mathrm{fl}}$ is negligible except for a few sporadic spots.

Because the anomalous resistivity used here activates only where $J > J_{\rm cr}$, both reconnection and Ohmic heating are confined to those high-current sites. Moreover, energy released by reconnection is partitioned into plasma motion and direct conversion to internal energy.  Consequently, maps of Joule heating, together with high-velocity outflows, can closely detect the locations of strong parallel electric fields, as shown by \cite{2025A&A...695A..40C} and \cite{reid2020determining}. 

A possible generalization of Eq. \ref{eq:clustering_2} then could be:
\begin{equation} 
        \mathbb{B} \, : \,
    E_{\parallel} \ne 0. 
    \label{eq:clustering_3}
\end{equation} 
where,     
\begin{equation} 
E_{\parallel} = \frac{\vec E \cdot \vec B}{|\vec B|} = \frac{J_{\parallel B}}{\sigma}. 
\end{equation} 

A simpler indicator often employed in uniform-resistivity models is the field-line integral of the parallel current $J_{\parallel}$ \citep[e.g.][]{wilmot2009magnetic}, used as a proxy for $\int_{\mathrm{fl}} E_{\parallel}\,ds$ \citep{schindler1988general, hesse1988theoretical}.  When resistivity is spatially varying or anomalous, however, the correspondence between current and electric field can break down \citep{reid2020determining}. \cite{reid2020determining} demonstrated instead that the field-line-integrated $E_{\parallel}$ and the integrated Ohmic heating share the same highly fragmented, spatially dispersed character, confirming that the dissipation and the reconnection electric field are essentially co-spatial in such models.

Sites of reconnection are often associated with narrow current sheets. In the presence of a strong guide field, formation of current sheets may be inhibited by the magnetic tension. On the other hand, when external conditions are favorable, their thickness is observed to decrease in time until their dissipation becomes strong enough (or it approaches the grid resolution, \citealt{rappazzo2013current}).
Possible candidates for dissipating current sheets can be therefore selected looking at their width:
\begin{equation} 
        \mathbb{B} \, : \,  
     |\vec B|/|\vec J| < \lambda_{tr}.
    \label{eq:clustering_4}
\end{equation} 

\begin{figure*}[p]
   \centering
\includegraphics[width=\hsize]{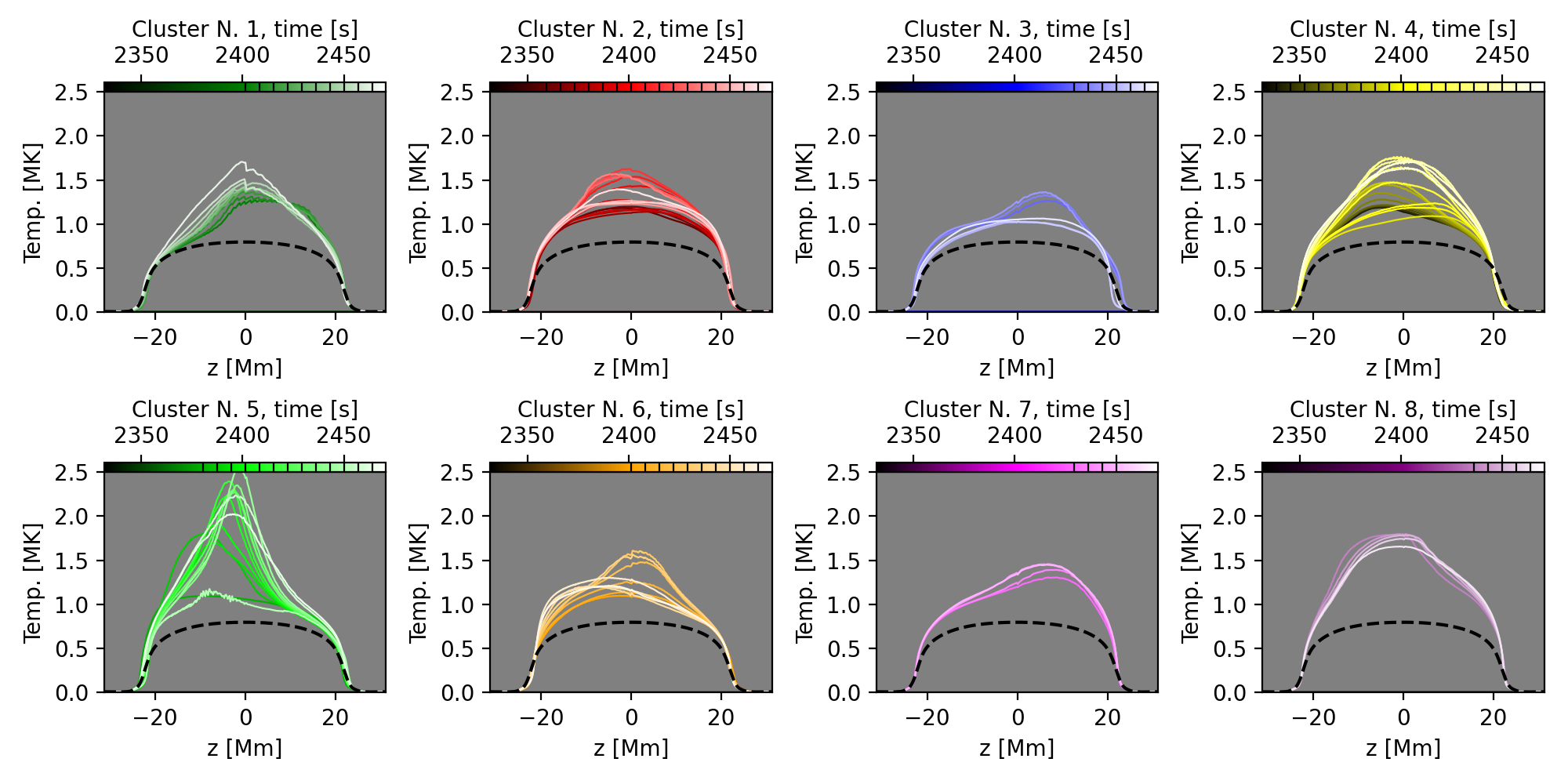}
  \caption{Field lines averaged values of temperature along the loop length ($z = 0$ is the apex) for the clusters n. 1 to 8 shown in Fig. \ref{Fig:paper_2_composition} at the different temporal stages pinned, in each colour-bar, by black ticks. The black, dashed lines show the spatially and temporally averaged background temperature along $\hat z$.} 
  \label{Fig:paper_2_temp_lines}
\end{figure*}

\section{Further examples of spatial and temporal characteristics of reconnection outflows}
\label{sec:appendix1}

\begin{figure*}[h!]
   \centering
\includegraphics[width=\hsize]{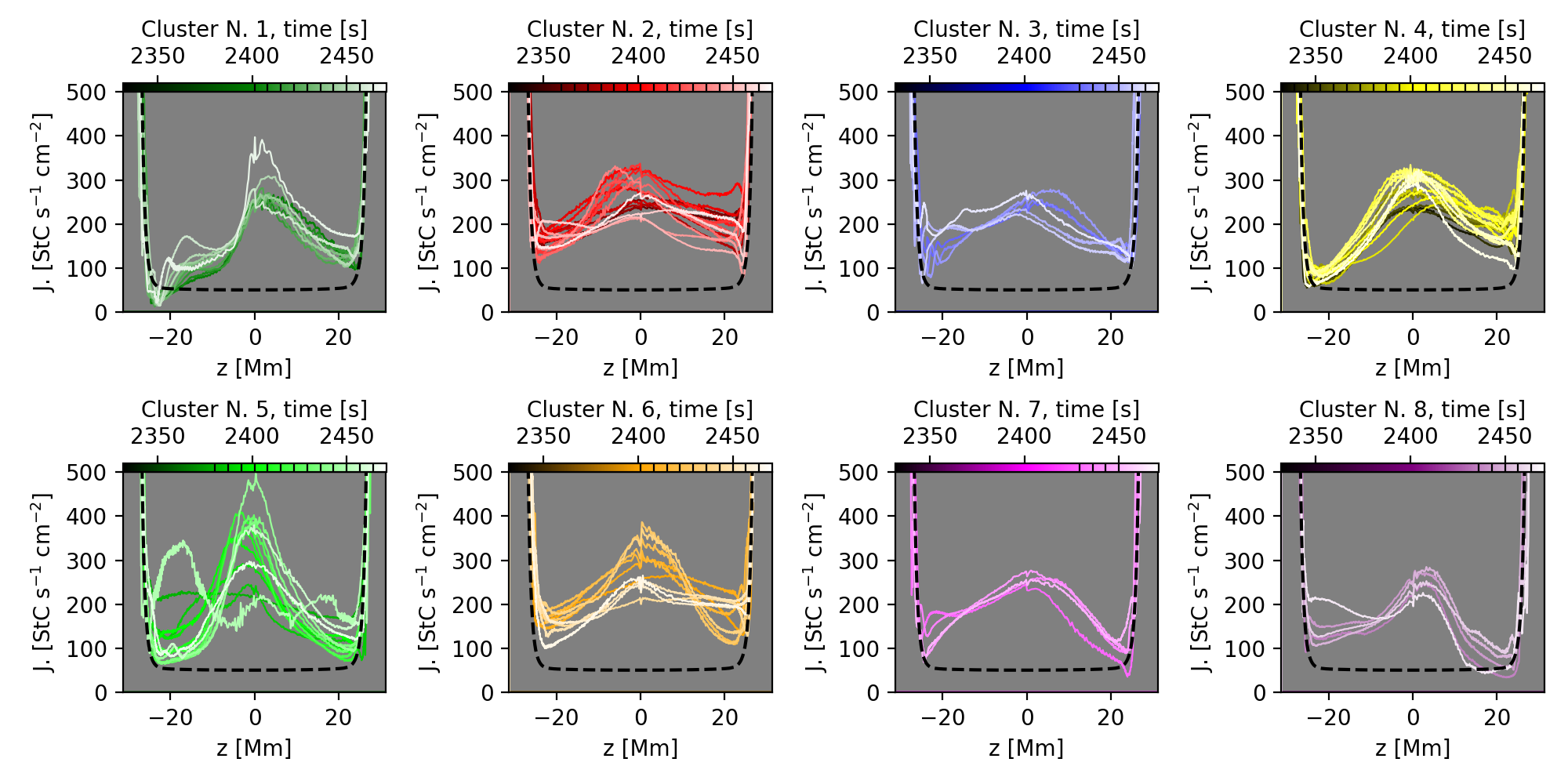}
  \caption{Field lines averaged values of current density along the loop length ($z = 0$ is the apex) for the clusters n. 1 to 8 shown in Fig. \ref{Fig:paper_2_composition} at the different temporal stages pinned, in each colour-bar, by black ticks. The black, dashed lines show the spatially and temporally averaged background current density along $\hat z$.} 
  \label{Fig:paper_2_current_lines}
\end{figure*}

\begin{figure*}[h!]
   \centering
\includegraphics[width=\hsize]{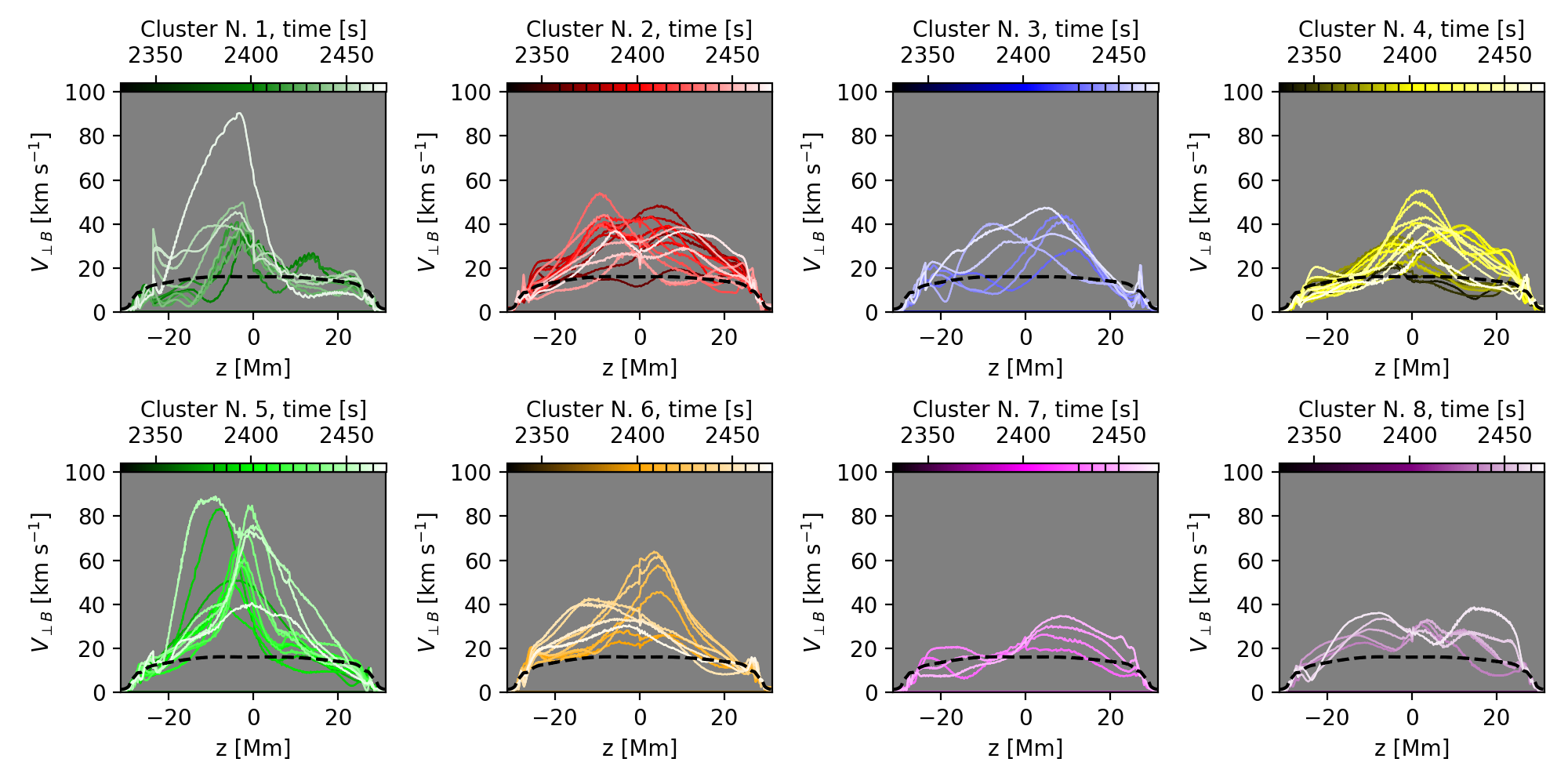}
  \caption{Field lines averaged values of the velocity along the loop length ($z = 0$ is the apex) for the clusters n. 1 to 8 shown in Fig. \ref{Fig:paper_2_composition} at the different temporal stages pinned, in each colour-bar, by black ticks. The black, dashed lines show the spatially and temporally averaged background velocity along $\hat z$.} 
  \label{Fig:paper_2_vel_lines}
\end{figure*}

\begin{figure*}[h!]
   \centering
\includegraphics[width=\hsize]{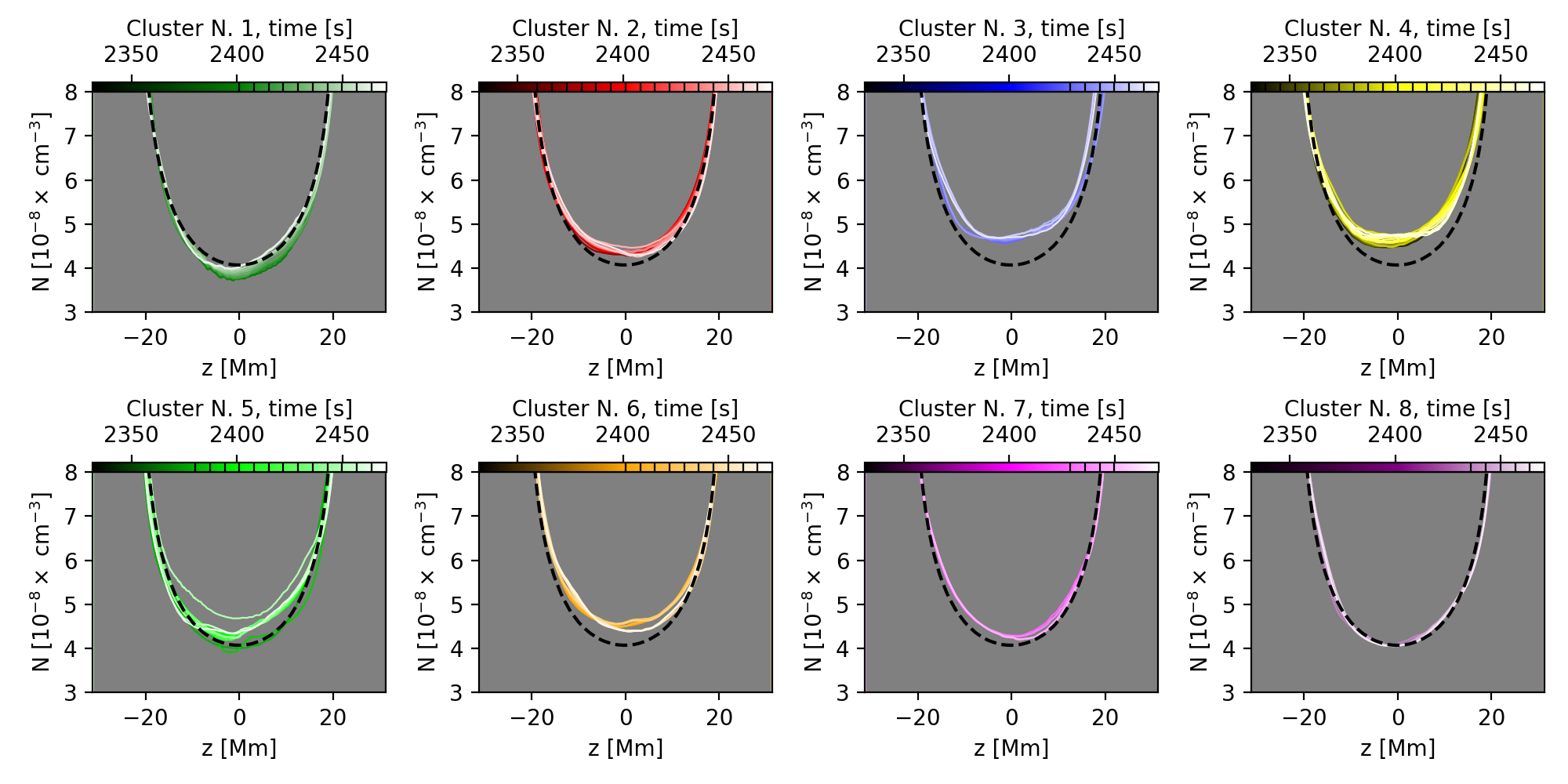}
  \caption{Field lines averaged values of plamsa density along the loop length ($z = 0$ is the apex) for the clusters n. 1 to 8 shown in Fig. \ref{Fig:paper_2_composition} at the different temporal stages pinned, in each colour-bar, by black ticks. The black, dashed lines show the spatially and temporally averaged background plasma density along $\hat z$.} 
  \label{Fig:paper_2_dens_lines}
\end{figure*}

Figures \ref{Fig:paper_2_temp_lines}-\ref{Fig:paper_2_dens_lines} extend the analysis of Fig. \ref{Fig:paper_2_composition} to all eight reconnection outflows identified at $t \sim 2400$ s.  For every event and at every stored snapshot we traced a bundle of field lines that intersect the local dissipation core (region $\mathbb{B}$; see Sect. 2 for the tracing procedure).  Along each line we sampled temperature, flow speed, current density, and mass density; the coloured curves in the figure show the values averaged over the ensemble of lines and plotted against arc length measured from one chromospheric foot-point to the other (loop apex at $s=0$).

A clear pattern emerges:
\begin{itemize}
    \item Temperature and current density peak sharply in the vicinity of the apex, marking the compact reconnection region.  Both quantities rise rapidly within a few seconds of the jet onset and decay on a comparable timescale, returning toward the background profile once the local current sheet has dissipated.

    \item Velocity is likewise concentrated near the apex at early times but its profile broadens as the outflow propagates, filling tens of Mm along the guide field.  Peak speeds remain in the $60\text{–}150$ km s$^{-1}$ range, consistent with the statistics in Fig. 6.

    \item Density displays a more gradual evolution.  During the impulsive phase it is nearly flat, reflecting the evacuation of plasma by the bidirectional jet.  As the event subsides, a modest density enhancement develops in the upper half of the loop, most likely the signature of chromospheric evaporation driven by the deposited heat.
\end{itemize}

Although individual events differ in magnitude and duration, the same qualitative behaviour recurs in all eight cases, reinforcing the interpretation that small-angle reconnection at the loop apex produces hot, fast, elongated jets whose thermodynamic aftermath includes gentle upflows from the chromosphere.

\bibliography{sample701}{}
\bibliographystyle{aasjournalv7}

\end{document}